\documentclass[letterpaper,aps,prd,preprint,showpacs,nofootinbib,superscriptaddress]{revtex4}
\pdfoutput=1
\usepackage{epsf}
\usepackage{epsfig}
\usepackage{graphics}
\usepackage{graphicx}
\usepackage{amssymb}

\newcommand{\nc}{\newcommand}
\nc{\postscript}[2]{\setlength{\epsfxsize}{#2\hsize}\centerline{\epsfbox{#1}}}

\nc{\beq}{\begin{equation}}   \nc{\eeq}{\end{equation}}
\nc{\bea}{\begin{eqnarray}}   \nc{\eea}{\end{eqnarray}}
\nc{\baa}{\begin{array}}      \nc{\eaa}{\end{array}}
\nc{\bit}{\begin{itemize}}    \nc{\eit}{\end{itemize}}
\nc{\ben}{\begin{enumerate}}  \nc{\een}{\end{enumerate}}
\nc{\bce}{\begin{center}}     \nc{\ece}{\end{center}}

\nc{\non}{\nonumber}

\begin{document}

\title{\begin{flushright}

\mbox{\normalsize \rm UMD-PP-09-053}\\\vspace{-10pt}
\mbox{\normalsize \rm UQAM-PHE-0901}
       \end{flushright}
\bf Patterns in the Fermion Mixing Matrix,\\ a bottom-up approach}

\author{Gilles Couture\footnote{couture.gilles@uqam.ca}}
\affiliation{ Groupe de Physique Th\'eorique des Particules,
D\'epartement des Sciences de la Terre et de L'Atmosph\`ere,
Universit\'e du Qu\'ebec \`a Montr\'eal, Case Postale 8888,
Succ. Centre-Ville, Montr\'eal, Qu\'ebec, Canada, H3C 3P8.
}

\author{Cherif Hamzaoui\footnote{hamzaoui.cherif@uqam.ca}}
\affiliation{ Groupe de Physique Th\'eorique des Particules,
D\'epartement des Sciences de la Terre et de L'Atmosph\`ere,
Universit\'e du Qu\'ebec \`a Montr\'eal, Case Postale 8888, Succ. Centre-Ville,
 Montr\'eal, Qu\'ebec, Canada, H3C 3P8.
}

\author{Steven S.~Y.~Lu\footnote{lu.steven@uqam.ca}}
\affiliation{ D\'epartement de Math\'ematiques,\\
Universit\'e du Qu\'ebec \`a Montr\'eal, Case Postale 8888, Succ. Centre-Ville,
 Montr\'eal, Qu\'ebec, Canada, H3C 3P8.}

\author{Manuel Toharia\footnote{mtoharia@umd.edu}}
\affiliation{Maryland Center for Fundamental Physics, \\ Department of
  Physics, University of Maryland, \\ College Park, MD 20742, USA.}

\begin{abstract}
We first obtain the most general and compact parametrization of the unitary transformation
diagonalizing any $3\times3$ hermitian matrix $H$, as a function of its elements
and eigenvalues. We then study a special class of fermion mass
matrices, defined by the requirement that all of the diagonalizing
unitary matrices (in the up, down, charged lepton and
neutrino sectors) contain at least one mixing angle much
smaller than the other two.
Our new parametrization allows us to quickly extract information
on the patterns and predictions emerging from this scheme. In
particular we find that the phase difference between two elements of
the two mass matrices (of the sector in question) controls the generic size
of one of the observable fermion mixing angles:
i.e. just fixing that particular phase difference will ``predict'' the generic
value of one of the mixing angles, irrespective of the value of
anything else.  

\end{abstract}

\maketitle

\section{Introduction}

In the absence of flavor symmetries, the Yukawa couplings between
the Standard Model (SM) fermions and the Higgs
field are in general complex arbitrary matrices which, after
Electroweak Symmetry Breaking (EWSB), become the mass
matrices of the quarks and charged leptons. In the case of
neutrinos, the mass matrix will be in general complex symmetric.
All these matrices contain more parameters than physical
observables and an explicit computation of these
observables (fermion masses and mixings) in terms of the original
matrix elements can be quite cumbersome in general. Indeed this
would require us to solve a $3\times 3$ eigenvalue problem for
each fermion matrix, and then compose the unitary transformations
(formed with the calculated eigenvectors) of the Up and Down quark
sectors and then also of the charged lepton and neutrino sectors
\footnote{Assuming three families of neutrinos, although
additional neutrino flavors are possible (sterile neutrinos).}.
We observe however that it may be useful to address the question
not as an eigenvalue problem, but as an eigenvector problem,
treating the eigenvalues as input parameters and not as output.
The first reason for this is that except for the neutrino sector,
all the mass eigenvalues are quite well known. But the main point
we make is that by keeping explicitly the mass eigenvalues as
input parameters, the eigenvector solutions of each mass matrix
become surprisingly simple and can be written as compact functions of
both the mass matrix elements and the eigenvalues (the fermion masses).
Such a parametrization of the mixing matrices, directly in terms
of the original mass parameters and the fermion masses might prove
to be useful in the studies aiming to explain the observed flavor
structure of the SM by way of symmetries or textures or patterns
\cite{Fritzsch1,Brokensymmetry,RRR,Textures} at the level of the fermion  
mass matrices.

It is true, though, that to proceed we need to work with hermitian
matrices, but it is always possible to render the quark mass matrices
hermitian in the Standard Model without loss of generality
\cite{Frampton,BrancoMota}.
The procedure to obtain hermitian matrices is quite standard, and it
involves either working with the hermitian matrix
$H_1=GG^\dagger$, where $G$ is the original fermion mass matrix, or using its
polar decomposition, i.e. solving $G = H_{{}_2} Q_{{}_G}$, where
$Q_{{}_G}$ is a unitary matrix converting the general complex
matrix $G$ into a positive semi-definite hermitian
matrix $H_{{}_2}$ (if $G$ is invertible, with distinct non-vanishing
eigenvalues, then $H_{{}_2}$ is positive definite and therefore $Q_{{}_G}$ is unique).

We will first present our parametrization (in fact 9 different types)
for the unitary matrix $W$ which diagonalizes a general hermitian matrix
$H$ in the most compact way possible.
Then, to start taking advantage of it, we then propose a
simple and mildly constraining ansatz for the flavor 
structure of the SM fermion sector.
It assumes that the unitary transformations diagonalizing $H_u$ and $H_d$ can each be decomposed
as only two rotations, instead of three. The idea is to assume that
the third rotation angle is zero (or much smaller than the other
two) and therefore one of the entries of each transformation matrix
$W_u$ and $W_d$ will be zero or close to zero. We call this setup the two-angle ansatz and we
will concentrate in only one of the many possible cases.
Our parametrization allows us to quickly obtain very simple dependences of the fermion
mixing matrices $V_{CKM}$ and $V_{PMNS}$  \cite{Cabibbo} in terms of the masses and
the original mass matrix elements.
We can thus study easily the interesting properties of this
type of ansatz as well as the consequences it has in the original mass
matrices, in both the quark and the lepton sectors. A particularly
interesting observation is that the specific value of some elements of
$H_u$ and $H_d$ has no effect (or very mild effect) on the observed
values of masses and mixings.

\section{Mixing matrix parametrization}

As explained before, we are going to concentrate on
hermitian matrices with the assumption that the fermion mass matrices
are either hermitian or that one can construct a hermitian matrix out of
them. All the results presented in this work are valid for general
hermitian matrices, but for simplicity we will only consider the case of
positive definite hermitian matrices. Let $H$ be one such matrix: 
\begin{eqnarray}
H=\pmatrix{ \gamma   & x   & g \cr x^* & \alpha   & b \cr g^* &
b^* & a \cr },
\end{eqnarray}
with eigenvalues $\lambda_1,\lambda_2$ and $\lambda_3$. A compact
parametrization of the unitary matrix $W$ which diagonalizes it,
is \footnote{There are 9 different ways of  parametrizing it,
  depending on the choice of which diagonal mass
  parameter ($\gamma,\ \alpha\ $ or $\ a\ $) is explicitly absent in
  each vector column (see Appendix A for details). In the parametrization shown here, the
  rotation matrix $W$ has the correct limit when the off-diagonal
  entries in the original mass matrices $H$ are set to zero, avoiding
  (apparent) divergences in this limit.}
\begin{equation}
\hspace{-.3cm}
W= \left(\begin{array}{ccc}
\frac{\displaystyle(\alpha-\lambda_1)(a-\lambda_1)-|b|^2}{\displaystyle N_1}
& \frac{\displaystyle gb^*-x(a-\lambda_2)}{\displaystyle N_2}
& \frac{\displaystyle \displaystyle xb -g(\alpha-\lambda_3)}{\displaystyle N_3}\\
\frac{\displaystyle g^*b-x^*(a-\lambda_1)}{\displaystyle N_1}
& \frac{\displaystyle (\gamma-\lambda_2)(a-\lambda_2)-|g|^2}{\displaystyle N_2}
& \frac{\displaystyle x^*g-b(\gamma-\lambda_3)}{\displaystyle N_3}\\
\frac{\displaystyle x^*b^*-g^*(\alpha-\lambda_1)}{\displaystyle N_1}
& \frac{\displaystyle xg^*-b^*(\gamma-\lambda_2)}{\displaystyle N_2}
& \frac{\displaystyle (\gamma-\lambda_3)(\alpha-\lambda_3)-|x|^2}{\displaystyle N_3}
\end{array}\right)\label{Wparametrization}
\end{equation}
After some algebra the normalization parameters are found to have the
simple form 
\begin{eqnarray}
N_1^2&=&(\lambda_3-\lambda_1)(\lambda_2-\lambda_1) \left[(\alpha-\lambda_1)(a-\lambda_1)-|b|^2\right],    \\
N_2^2&=&(\lambda_3-\lambda_2)(\lambda_2-\lambda_1) \left[(a-\lambda_2)(\lambda_2-\gamma)+|g|^2\right],    \\
N_3^2&=&(\lambda_3-\lambda_2)(\lambda_3-\lambda_1) \left[(\lambda_3-\gamma)(\lambda_3-\alpha)-|x|^2\right]
\end{eqnarray}
The surprisingly simple and compact form of this parametrization might
make it suitable to treat flavor models keeping always an explicit
dependence on all the matrix elements of the hermitian mass matrices.
Of course, if the three eigenvalues
$\lambda_1$, $\lambda_2$ and $\lambda_3$ are fixed, there must be three
constraint equations on the elements of the matrix $H$. These
equations are found from the three invariants $Tr(H)$, $Tr(H^2)$ and $Det(H)$: 
\begin{eqnarray}
Tr(H)&=&a+\alpha+\gamma=\lambda_1+\lambda_2+\lambda_3,  \\
Tr(H^2)&=&2(|x|^2+|b|^2+|g|^2)+a^2+\alpha^2+\gamma^2=\lambda_1^2+\lambda_2^2+\lambda_3^2  \\
Det(H)&=&\gamma(\alpha a -|b|^2)-a|x|^2 -\alpha |g|^2 + 2Re(bxg^*)= \lambda_1\lambda_2\lambda_3
\end{eqnarray}
By choosing $g$ as an independent variable, it
is possible to rewrite these constraint relations on the rest of
variables as\footnote{The same type of relations
  can be written when choosing $x$ or $b$ as the independent variable.}
\begin{eqnarray}
\alpha&=&\lambda_1+\lambda_2+\lambda_3-a-\gamma,\\
|x|^2&=&\frac{(\gamma-\lambda_1)(\lambda_2-\gamma)(\lambda_3-\gamma)-|g|^2(\alpha-\gamma)+2 Re(bxg^*)}{(a-\gamma)}\\
|b|^2&=&\frac{(a-\lambda_1)(a-\lambda_2)(\lambda_3-a)+|g|^2(\alpha-a)-2 Re(bxg^*)  }{(a-\gamma)}
\label{bxgconstraints}
\end{eqnarray}
The interesting thing of this notation is that the constraint formulae
on $x$ and $b$ actually become algebraic solutions for both $x$ and
$b$ when the term $Re(bxg^*)$ vanishes identically. In particular,
this is the case when one deals with mass matrices with texture zeroes
in the off-diagonal elements.

\section{Flavor in the two-angle ansatz}

Equipped with an exact and simple parametrization of the fermion
mixing matrix in both Up and Down sectors (or charged lepton and
neutrino sectors), we look for economical patterns among the mixing
matrices by following a bottom-up approach in the hope that it might
be complementary to more top-down approaches such as imposing flavor
symmetries or texture zeroes mass matrices (see \cite{Fritzsch1,Brokensymmetry,RRR,Textures} as well as the
probably incomplete surveys of \cite{Flavor,mutau}).
One avenue is to find a similar ansatz for the flavor structure
of the mixing matrices in both quark and leptonic sectors.
Such possibility exists in the sense that in both sectors, the mixing
elements $(V_{CKM})_{13}$ and $(V_{PMNS})_{13}$ are known to be small. This feature is very
interesting and it is known to lead to a simple parametrization of the
mixing matrix. Note that in the limit $V_{13}=0$ we have two
additional conditions, since $V_{13}$ is a complex number. This means
that in this limit, the mixing matrix will have only two independent
parameters instead of four.
We can choose these four independent parameters to be $|V_{12}|$,
$|V_{23}|$, $|V_{13}|$ and $|V_{21}|$ \cite{CherifModulii}.
In the limit of $V_{13}=0$, we have the extra constraint
$|V_{21}|=|V_{12}|\sqrt{1-|V_{23}|^2}$. This means that the whole
mixing matrix is described by $|V_{12}|$ and $|V_{23}|$.
Note also that in this limit, there is no ${\cal CP}$ violation \`a
la Dirac in both quark and leptonic sectors.

In what follows, the
subscript $0$ stands for the values of the mixing matrix elements in
the limit $V_{13}=0$. Since $V_{13}$ is known to be very small, we
believe that the zeroes values of the mixing matrix elements are not
far from their measured values. For the quarks, we have:
\begin{equation}
V_{CKM}^0= \left (
\begin{array}{ccc}
\sqrt{1-|V^0_{us}|^2} & |V^0_{us}| & 0 \\
-|V^0_{us}|\sqrt{1-|V^0_{cb}|^2} & \sqrt{(1-|V^0_{us}|^2)(1-|V^0_{cb}|^2)}  & |V^0_{cb}|  \\
|V^0_{us}| |V^0_{cb}| & -|V^0_{cb}| \sqrt{1-|V^0_{us}|^2}  &
\sqrt{1-|V^0_{cb}|^2}
\end{array} \right )
\end{equation}
Note that this zero-order $V_{CKM}^0$ can be decomposed as a product of two
rotations, namely one is purely the Cabbibo angle and the other
one is purely made out of beauty namely $|V_{cb}|$:
\begin{equation}
V_{CKM}^0 = V_{B}V_{C}
\end{equation}
with
\begin{equation}
\hspace{-.2cm}V_{B}\!=\!\pmatrix{1 & 0 & 0 \cr 0 & \sqrt{1-|V^0_{cb}|^2} &
|V^0_{cb}| \cr 0 & -|V^0_{cb}| & \sqrt{1-|V^0_{cb}|^2} \cr }
\ \ {\rm and}\ \
V_{C}\!=\!\pmatrix{ \sqrt{1-|V^0_{us}|^2} & |V^0_{us}| & 0 \cr
-|V^0_{us}| & \sqrt{1-|V^0_{us}|^2} & 0 \cr 0 & 0 & 1 \cr }
\end{equation}
In the leptonic sector we have:
\begin{equation}
V_{PMNS}^0= \left (
\begin{array}{ccc}
\sqrt{1-|V^0_{e2}|^2} & |V^0_{e2}| & 0 \\
-|V^0_{e2}|\sqrt{1-|V^0_{\mu 3}|^2} & \sqrt{(1-|V^0_{e2}|^2)(1-|V^0_{\mu 3}|^2)} & |V^0_{\mu 3}|  \\
|V^0_{e2}| |V^0_{\mu 3}| & -|V^0_{\mu 3}| \sqrt{1-|V^0_{e2}|^2}
& \sqrt{1-|V^0_{\mu 3}|^2}
\end{array} \right )P
\end{equation}
which can be decomposed also as a product of two
rotations, namely one is purely solar and the other one is purely
atmospheric:
\begin{equation}
V_{PMNS}^0 = V_{atm}V_{sol}P
\end{equation}
with,
\begin{equation}
\hspace{-.3cm}V_{atm}\!=\!\pmatrix{1 & 0 & 0 \cr 0 & \sqrt{1-|V^0_{\mu3}|^2} &
|V^0_{\mu3}| \cr 0 & -|V^0_{\mu3}| & \sqrt{1-|V^0_{\mu3}|^2} \cr }
\ \ {\rm and}\ \
V_{sol}\!=\!\pmatrix{ \sqrt{1-|V^0_{e2}|^2} & |V^0_{e2}| & 0 \cr -|V^0_{e2}|
& \sqrt{1-|V^0_{e2}|^2} & 0 \cr 0 & 0 & 1 \cr }
\end{equation}
The diagonal phase matrix $P$ which contains the Majorana phases
is defined as:
\begin{eqnarray}
P = \pmatrix{ 1 & 0 & 0 \cr 0 & e^{i\eta} & 0 \cr 0 & 0 & e^{i\xi}\cr}
\end{eqnarray}
This structure for the physical fermion mixing matrices, decomposed
mainly into just two rotations is quite suggestive and we will use
this observation as the starting point of our analysis.
In this context we ask ourselves how many economical possibilities one
has to restore minimally and fully the mixing elements and ${\cal CP}$ violation.
We start with the structure
\begin{eqnarray}
W_u^0=\pmatrix{ 1 & 0 & 0 \cr 0 & X & X \cr 0 &
X & X \cr },\phantom{pp}
W_d^0=\pmatrix{ X  & X & 0 \cr X & X & 0 \cr 0 &
0 & 1 \cr },
\end{eqnarray}
and then to be as general as possible while still keeping the original
motivation we consider all mixing patterns emerging from the original
structure, but with only one zero in each mixing matrix. Now, we
establish all the corrected mixing patterns for both Up and Down quark
sectors:
\begin{eqnarray}
\hspace{-1cm}
W_u^C \equiv \pmatrix{ X  & 0 & Cor \cr Cor & X & X \cr Cor &
X & X \cr }
,
\pmatrix{ X  & Cor & 0 \cr Cor & X & X \cr Cor &
X & X \cr }
,
\pmatrix{ X  & Cor & Cor \cr 0 & X & X \cr Cor &
X & X \cr }
,
\pmatrix{ X & Cor & Cor \cr Cor & X & X \cr 0 &
X & X \cr },\ \ \
\end{eqnarray}
\begin{eqnarray}
\hspace{-1cm}
W_d^C \equiv \pmatrix{ X  & X & 0 \cr X & X & Cor \cr Cor &
Cor & X \cr },
\pmatrix{ X  & X & Cor \cr X & X & 0 \cr Cor &
Cor & X \cr },
\pmatrix{ X & X & Cor \cr X & X & Cor \cr 0
& Cor & X \cr },
\pmatrix{ X  & X & Cor \cr X & X & Cor \cr Cor &
0 & X \cr },\ \ \
\end{eqnarray}
where $X$ stands for a non-zero value and $Cor$ for a corrected originally zero mixing matrix element.
These mixing matrices with one texture zero in them can be decomposed
themselves into only two rotations instead of three. We call this the
two-angle ansatz, in which both the Up and Down quark sectors are
diagonalized by unitary transformations containing just two angles
(i.e. having one vanishing element).
There are obviously many possibilities for this ansatz but in particular 
there are 16 cases such that one can recover in a specific limit the
case $V_{13}=0$. In this work we are going to focus only on one
specific example of this type of ansatz, although a full case by case
study is underway.
We feel that the main features of this ansatz do reveal
themselves in the example studied here, and we prefer to continue elsewhere
a more systematic exploration.\\

{\bf Notation}\\
Before we proceed further, we will set up
our notation for all the matrices in both up and down quark sectors as
well as charged lepton and neutrino sectors.
We will define the Hermitian matrices $H_u$, $H_d$, $H_l$ and $H_\nu$ defined by:
\bea
H_u=\pmatrix{ \gamma   & x   & g \cr x^* & \alpha   & b \cr g^* &
b^* & a \cr }
,\phantom{pp}
 H_d=\pmatrix{ \rho  & y   & h \cr y^* &
 \beta   & f \cr h^*  & f^* & d \cr} .\label{quarksmass}
\eea
with eigenvalues $m_u,m_c,m_t,m_d,m_s$ and $m_b\ $
and
\bea
H_l=\pmatrix{ \gamma'   & x'   & g' \cr x'^* & \alpha'   & b' \cr g'^* &
b'^* & a' \cr }
,\phantom{pp}
 H_\nu=\pmatrix{ \rho'  & y'   & h' \cr y'^* &
 \beta'   & f' \cr h'^*  & f'^* & d' \cr} .
\label{leptonsmass}
\eea
with eigenvalues $m_e,m_\mu,m_\tau,\lambda_1,\lambda_2$ and $\lambda_3$.

In the text, we will also need to refer to the phases of the
off-diagonal terms, denoted as
\bea
&&\arg{(H_u)_{12}}=\delta_x,\hspace{.8cm}\arg{(H_u)_{13}}=\delta_g,\hspace{.8cm} \arg{(H_u)_{23}}=\delta_b,\\
&&\arg{(H_d)_{12}}=\delta_y,\hspace{.8cm}\arg{(H_d)_{13}}=\delta_h,\hspace{.8cm} \arg{(H_d)_{23}}=\delta_f,
\eea
and similarly for the leptons.
The matrices diagonalizing these mass matrices will be denoted
respectively as $W_u$, $W_d$, $W_l$ and $W_\nu$.


\subsection{THE CASE (13-13): $\bf \ \Big({\rm i.e.}\  (W_u)_{13}=(W_d)_{13}=0\ \ {\rm and/or}\ \ (W_l)_{13}=(W_\nu)_{13}=0 \Big)$}


We will now consider the mass matrices from Eqs.~(\ref{quarksmass}) and
(\ref{leptonsmass}) with the extra constraints of $(W_u)_{13}=0$
and $(W_d)_{13}=0$ in the quark sector and $(W_l)_{13}=0$
and $(W_\nu)_{13}=0$ in the lepton sector.\footnote{
In the quark sector, this limit was shown to lead to acceptable
patterns in the limit $\rho=0$ and $\gamma=0$ \cite{BrancoRebelo} and similar limits were 
considered in the lepton sector in \cite{Antusch}}

\subsubsection{\bf The quark sector}
For example in the down quark sector imposing $(W_d)_{13}=0$
corresponds to the requirement $\ yf- h(\beta-m_b)=0 \ $.
After a short computation, we obtain simpler relations among the
elements of the mass matrix:
\begin{eqnarray}
|y|^2 &=&\frac{(\rho-m_d)(m_s-\rho)(m_b-\beta)}{(2m_b-\beta-d)}\nonumber\\
|h|^2 &=&\frac{(\rho-m_d)(m_s-\rho)(m_b-d)}{(2m_b-\beta-d)}\nonumber\\
|f|^2 &=& (m_b-\beta)(m_b-d) \nonumber\\
\delta_{y} -\delta_{h} + \delta_{f} &=& \pi
\end{eqnarray}
These relations will simplify significantly the general form of the Down quark mixing
matrix $W_d$ as well as the up quark mixing matrix $W_u$
(see Eq.~(\ref{Wparametrization}) and Appendix B for details).

It is then straightforward to show that the quark mixing matrix
takes the form \bea
V&=&W_u^\dagger W_d=P^\dagger_u\ V_{{}_{CKM}}\ P_d
\eea
where $P_u$ and $P_d$ are unphysical diagonal phase matrices given
below and
\bea
V_{{}_{CKM}}=\pmatrix{
c_{\gamma}c_\rho+s_{\rho}s_{\gamma} |C| e^{i\theta}
&-c_{\gamma}s_\rho+s_{\gamma}c_{\rho} |C| e^{i\theta}
&s_\gamma\ |S|
\cr
-s_{\gamma}c_\rho+c_{\gamma}s_{\rho} |C| e^{i\theta}
& s_{\gamma}s_{\rho}+c_{\gamma}c_{\rho} |C|e^{i\theta}
& c_\gamma |S|
\cr
-s_{\rho} |S|
&-c_{\rho}  |S|
&
|C| e^{-i\theta}
\cr }. \label{vckm}
\eea
The complex rotation parameters $C$, $S$, $c_\gamma$ and $c_\rho$ are given by
\bea C&=&
\frac{\left(\vphantom{\int_{\int}^{\int}}
  \sqrt{\displaystyle (m_t-a)(m_b-d)}\ e^{i(\delta_f-\delta_b)} +\sqrt{(m_t-\alpha)(m_b-\beta)}\right)
}{\sqrt{(2m_t-\alpha-a)(2m_b-\beta-d)}}\ \ \ (i.e.\ \ |C|
\equiv\ |V_{tb}|)\ \ \ \label{Ceq}\\
S&=&e^{i \delta_b} \frac{\left(\vphantom{\int_{\int}^{\int}}
  \sqrt{\displaystyle (m_t-a)(m_b-\beta)}- e^{i(\delta_f-\delta_b)} \sqrt{(m_t-\alpha)(m_b-d)}\right)
}{\sqrt{(2m_t-\alpha-a)(2m_b-\beta-d)}} \label{Seq}\\
c_{\gamma}&=& \sqrt{\frac{m_c-\gamma}{m_c-m_u}}\label{cgammaeq}\\
c_\rho &=&\sqrt{\frac{m_s-\rho}{m_s-m_d}}\label{crhoeq}
\eea
Note that $s_i=\sqrt{1-c_i^2}$ and that $|C|^2+|S|^2=1$.
The CP phase $\theta$ and the two unphysical phase matrices $P_u$ and
$P_d$ are given respectively by $\theta=\delta_x-\delta_y-\delta_C\ $,
$\ P_u=diag(1,e^{i\delta_x},e^{i(\delta_y+\delta_S)})$
and $P_d=diag(1,e^{i\delta_y},e^{i(\delta_x+\delta_S)})$, with $\delta_i\equiv\arg{(i)}$.

The previous form of the mixing matrix implies the following exact 
relations for the quark sector
\bea
\frac{|V_{ub}|}{|V_{cb}|}=\sqrt{\frac{\gamma-m_u}{m_c-\gamma}},\hspace{.7cm}
\frac{|V_{td}|}{|V_{ts}|}=\sqrt{\frac{\rho-m_d}{m_s-\rho}}.
\label{VubVcb}
\eea
and 
\bea
\frac{|V_{ub}|}{|V_{td}|}=\sqrt{\frac{(\gamma-m_u)(m_s-m_d)}{(m_c-m_u)(\rho-m_d)}}, \hspace{.7cm}
\frac{|V_{cb}|}{|V_{ts}|}=\sqrt{\frac{(m_c-\gamma)(m_s-m_d)}{(m_c-m_u)(m_s-\rho)}},\label{VubVtd}
\eea
which really correspond to two independent constraints (i.e. the third
and fourth relations can be obtained using the first two along with
the unitary constraints).

As can be seen only two original mass matrix elements $\gamma$ (from
$H_u$) and $\rho$ (from $H_d$) appear explicitly in these last four
relations showing the first effect of the $(13-13)$ ansatz,
i.e. linking each of the previous ratios to a different quark mass
matrix, and in particular to the first diagonal elements of each mass
matrix. Once we fix these two elements to fit the experimental value of the
ratios given in Eqs.~(\ref{VubVcb}) and (\ref{VubVtd}), we will still have 6 free parameters (including 4 phases) to fit
the rest of the data (i.e we need to fit two more scalar observables from
  experiment, which for example can be taken to be the absolute
  value of $|V_{tb}|$ and the value of the CP phase of the fermion
  mixing matrix). 
We can choose to use $a,d$ and the phases of $x,g,f$
and $b$ as the free parameters. It is also useful to define the phase combinations:
\bea
\Delta_{23}\ =\ \arg{(H_d)_{23}}-\arg{(H_u)_{23}}&=&\delta_f-\delta_b\ ,\\
\Delta_{12}\ =\ \arg{(H_d)_{12}}-\arg{(H_u)_{12}}&=&\delta_y-\delta_x\ \\
\Delta_{13}\ =\ \arg{(H_d)_{13}}-\arg{(H_u)_{13}}&=&\delta_h-\delta_g.
\eea
which have the constraint
\bea
\Delta_{23}\ + \Delta_{12}\ =\ \Delta_{13}.
\eea
due to the $(13-13)$ ansatz imposition.

 Of course, we have more than enough free parameters to fit the two
 remaining observables, but as we will shortly see, of the 4 free phases, only two phase differences
  can be relevant, and the 2 real parameters $a$ and $d$ turn out to
  be statistically irrelevant if the two phases are properly
  chosen. In other words, in most of the parameter space of the six
  free parameters needed to produce the two remaining physical observables,
  4 directions are more or less irrelevant, with two phase differences
  being the two parameters required to obtain a good experimental
  fit. This situation is somewhat surprising because the amount of ``useful" free parameters is
   less than the total number of free parameters. Let's see how this is
  played out.

\begin{figure}[t]
 \center
\includegraphics[width=7.2cm,height=8.cm]{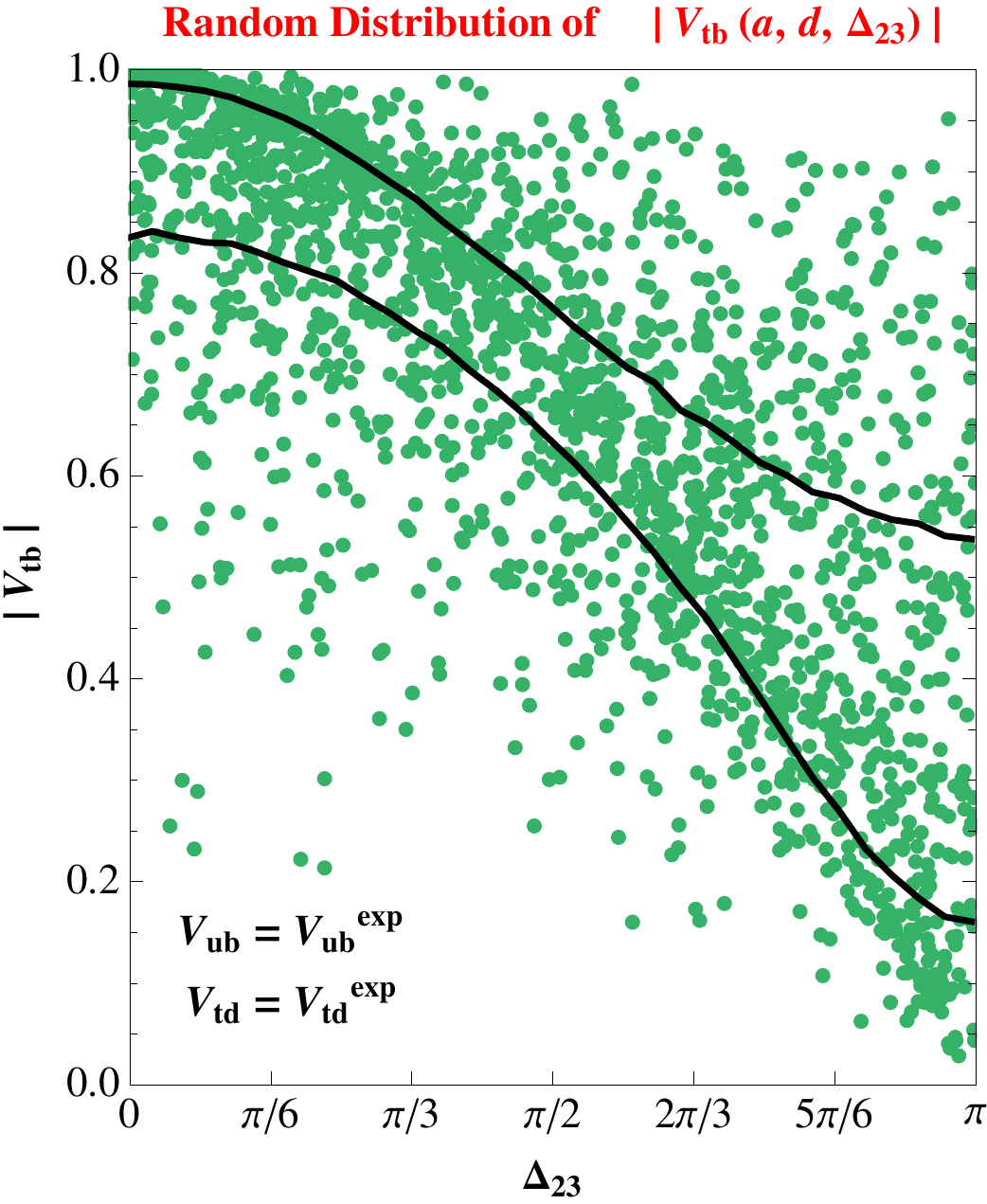}\hspace{1cm}
\includegraphics[width=7.2cm,height=8.cm]{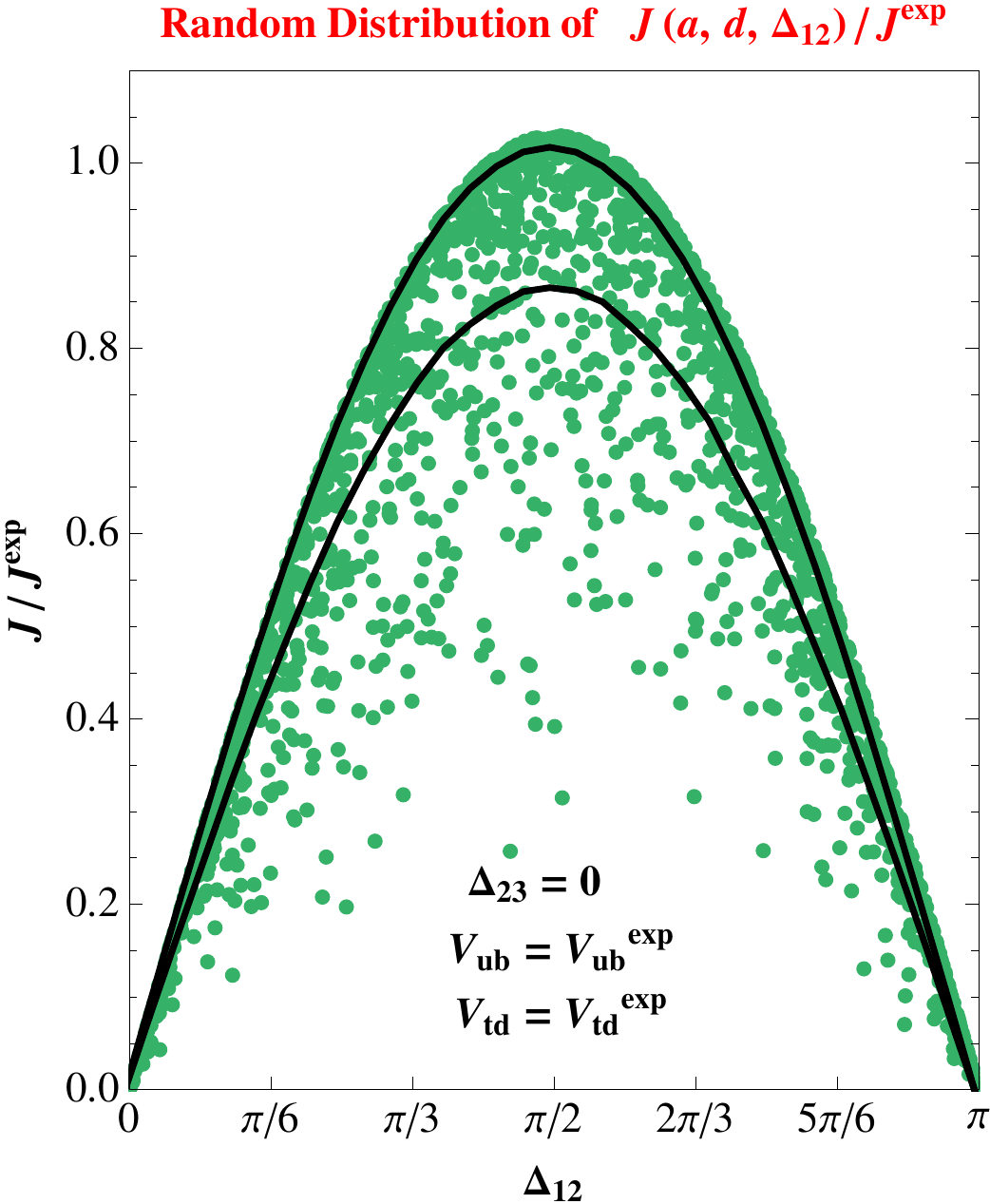}
 \caption{Distribution of $|V_{tb}(a,d,\Delta_{23})|$ with random $a$,
   $d$ and $\Delta_{23}$ with respect to $\Delta_{23}$
   (left panel), where $a$ and $d$ are two diagonal elements of the
   quark mass matrices $H_u$ and $H_d$, and $\Delta_{23}$ is the
   difference between the phases of the $H_{(23)}$ elements of these matrices. On the
   right panel, we present the
   distribution of the Jarlskog invariant $J(a,d,\Delta_{12})$ with
   random $a,d$ and $\Delta_{12}$, for $\Delta_{23}=0$. On the two
   panels, the black curves represent the $25\%$ and $75\%$
   quantiles of the distribution for fixed $\Delta_{23}$ (left) and
   $\Delta_{12}$ (right). In other words $50\%$ of the random points
   lie between the curves, with $25\%$ above it and $25\%$ below it.  }
\label{vtbdelta}
\vspace{.2cm}
 \end{figure}
Eq.~(\ref{VubVcb}) shows that $\gamma \equiv{\cal O}(m_u)$ and
$\rho\equiv{\cal O}(m_d)$ are required in order to obtain a good fit with
experimental data~\cite{PDG}. This has interesting implications
for $V_{tb}$ since it means that $a+\alpha = m_t+m_c +{\cal O}(m_u)$
and $d+\beta=m_b+m_s + {\cal O}(m_d)$ after using the trace identity of $H_u$ and
$H_d$.
We can therefore write
\bea
V_{tb}(a,d,{\Delta_{23}}) \simeq \frac{\sqrt{(m_t-a)(m_b-d)}\ e^{i
    {\Delta_{23}}}+\sqrt{(a-m_c)(d-m_s)}}{\sqrt{(m_t-m_c)(m_b-m_s)}}
\eea
It turns out that when the phase ${\Delta_{23}}$ is small (modulo $2\pi$),
statistically we find that
$|V_{tb}(a,d,0)| \sim 1$ for any randomly chosen value of
$a$ and $d$. In fact, the generic value of $|V_{tb}(a,d,{\Delta_{23}})|$ is very
much correlated with the value of ${\Delta_{23}}$, with little dependence
on the values of the other two variables, at least for small enough
${\Delta_{23}}$. This is shown in figure \ref{vtbdelta}, where we plot
the distribution of $|V_{tb}(a,d,{\Delta_{23}})|$ with respect to ${\Delta_{23}}$
for randomly chosen values of $a,d$ and $\Delta_{23}$. It is apparent
that there is a clear correlation between the value of the phase
${\Delta_{23}}$ and the value of $|V_{tb}|$.
The two black curves correspond to the $25\%$ and $75\%$ quantiles of
the distribution of $|V_{tb}|$ for a given value of ${\Delta_{23}}$ (i.e. $50\%$
of the randomly generated points lie between the two curves, with $25\%$
above them and $25\%$ below).

\begin{figure}[t]
 \center
 \includegraphics[width=9.cm,height=7.cm]{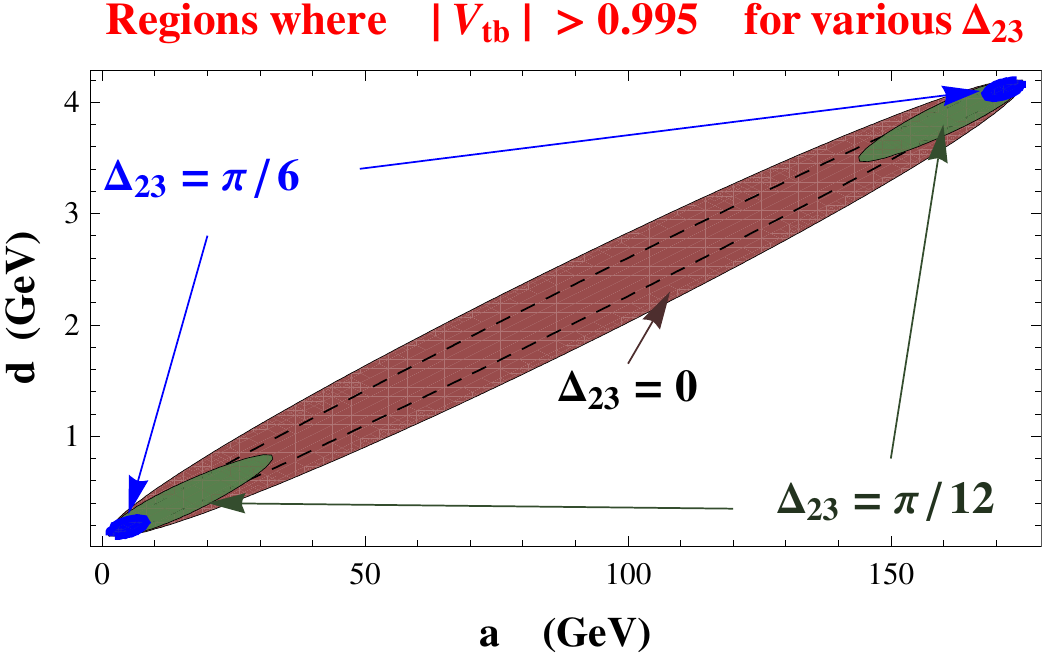}
  \caption{Regions in the plane $(a,d)$ where $|V_{tb}|>0.995$, for
    three different values of $\Delta_{23}$. The dashed curve is a contour of the experimental fit
    for $V_{tb}\simeq0.9991$ in the case $\Delta_{23}=0$. One sees
    quite clearly that as soon as $\Delta_{23}$ is increased, the
    parameter space favorable for large values of $V_{tb}$ shrinks dramatically.}
\label{adregion}
 \end{figure}

All this shows that the generic value of $V_{tb}$ in this ansatz
is actually governed by the specific value of the phase ${\Delta_{23}}$, with
very mild dependence on the other 2 parameters $a$ and $d$.
It is not clear though, that demanding ${\Delta_{23}}\sim 0$ will be
enough to fit the observed data in the quark sector since
$V^{fit}_{tb}\simeq 0.999$, i.e it is quite close to 1. The
random scan of Figure \ref{vtbdelta} does not show the distribution of $V_{tb}$
for $\Delta_{23} =0$, but once we fix $\Delta_{23}$
we just have two free parameters, $a$ and $d$. In Figure
$\ref{adregion}$ we show the complete allowed phase space for
$a$ and $d$, which are subject to the experimental constraints
$\ m_u<a<m_t\ $ and $\ m_d<d<m_b\ $. The different shaded regions are the points
where $V_{tb} > 0.995$, for three different values of the phase
difference $\Delta_{23}$. When this last one is zero, the region is quite
large, and it decreases very fast as the phase is increased. Because
the area of parameter space can be quite large we may still say that
the value of $|V_{tb}|$ consistent with experimental data could be
considered ``generic'' as long as the phase difference $\Delta_{23}$ is
vanishingly small.

Finally, one can also compute quite easily the Jarlskog invariant \cite{Jarlskogbook} in this
context. From Eq.~(\ref{vckm}), and using for example the definition $J={\rm
  Im}(V_{cb}V^*_{tb}V^*_{cd}V_{td})$, it is easy to see that it will have the form:
\bea
J&=& c_\rho c_\gamma\ s_\rho s_\gamma\ |C||S|^2 \sin{\theta} \label{j1313}
\eea
where $c_i$, $C$ and $S$ were given in Eqs.~(\ref{Ceq}), (\ref{Seq}),
(\ref{cgammaeq}) and (\ref{crhoeq}), and
where $\theta=\delta_x-\delta_y-\delta_C$ with $\delta_C=\arg{C}$.

Before analyzing in more detail the dependence on the original mass
matrix elements, it is interesting to relate the phase $\theta$ not just
to the Jarlskog invariant but also to the angles of the unitarity
triangle $\ \phi_1 = \beta\ $, $\ \phi_2 =\alpha\ $ and $\ \phi_3
 = \gamma $. These angles are defined as:
\bea
\phi_1 & =& \beta \
=\ \arg{(-\frac{V_{cd}V_{cb}^*}{V_{td}V_{tb}^*})}\ , \\
\phi_2 & =& \alpha \ 
=\ \arg{(-\frac{V_{td}V_{tb}^*}{V_{ud}V_{ub}^*})}\  , \\
\phi_3 & =& \gamma \ 
=\ \arg{(-\frac{V_{ud}V_{ub}^*}{V_{cd}V_{cb}^*})} 
\eea
and the relation between them and the Jarlskog J is:
\bea
\hspace{-.4cm}J\  =\   |V_{ub}||V_{td}||V_{tb}||V_{ud}|\ \sin{\alpha} 
\  =  \
|V_{td}||V_{cb}||V_{tb}||V_{cd}|\ \sin{\beta}
\  = \  
|V_{ub}||V_{cb}||V_{cd}||V_{ud}|\ \sin{\gamma}\label{Jalpha}
\eea
In the case of our ansatz, we can rewrite our J as
\bea
\hspace{-.4cm}J\ =\ |V_{ub}||V_{td}| |V_{tb}| c_\rho c_\gamma\ \sin{\theta}
\ =\  
|V_{td}||V_{cb}| |V_{tb}| c_\rho s_\gamma\ \sin{\theta} 
\ = \
|V_{ub}||V_{cb}||V_{tb}| c_\rho s_\rho \ \sin{\theta}\label{J1313}
\eea
For example, from the first identities of Eq.~(\ref{Jalpha}) and (\ref{J1313}) one
sees that we must have
\bea
\sin{\alpha}&=& \frac{c_\rho c_\gamma\ \sin{\theta}}{|V_{ud}|}
\eea
Since $c_\rho c_\gamma\ \simeq |V_{ud}|$, it follows from the above equation that
$\alpha \simeq \theta$. The experimental constraints on $\alpha$ are
such that $\alpha=(88^{+6}_{-5})^\circ$ \cite{PDG}, which basically means that
the phase $\theta$ is constrainted to be $\theta\simeq \pi/2$.

A more revealing way to see what this means for the original elements of the
mass matrices, we can actually rewrite the Jarlskog in a more explicit
way as 
\bea
J&=& |V_{ub}||V_{td}| c_\rho c_\gamma\   {\frac{\left(\vphantom{\int_{\int}^{\int}}
  \sqrt{\scriptstyle  (m_t-a)(m_b-d)}\ \sin{\scriptstyle (\Delta_{23}+\Delta_{12})} +
  \sqrt{\scriptstyle (m_t-\alpha)(m_b-\beta)}\ \sin{\scriptstyle
    \Delta_{12}}\right)}{\sqrt{\scriptstyle
      (2m_t-\alpha-a)(2m_b-\beta-d)}}}\\
&\simeq& |V_{ub}||V_{td}| c_\rho
c_\gamma\   {\frac{\left(\vphantom{\int_{\int}^{\int}}  \sqrt{\scriptstyle  (m_t-a)(m_b-d)}\ \sin{\scriptstyle (\Delta_{23}+\Delta_{12})} +
  \sqrt{\scriptstyle (a-m_c)(d-m_s)}\ \sin{\scriptstyle
    \Delta_{12}}\right)}{\sqrt{\scriptstyle
      (m_t-m_c)(m_b-m_s)}}}\label{Jfunct}
\eea
where the approximation of the second line comes from assuming that $\ a+\alpha = m_t+m_c
+{\cal O}(m_u)\ $ and $\ d+\beta=m_b+m_s + {\cal O}(m_d)$ .

We showed earlier that the imposition of $\Delta_{23}\equiv(\delta_f-\delta_b)\sim
0\ $ gives a nice statistical reason for the large value of
$|V_{tb}|$. With that extra condition, one actually has $\Delta_{12}\sim
\theta$, which then means that $\Delta_{12}\sim \pi/2$ due to experimental bounds.
The experimental best fit value of the Jarlskog invariant is
$|J_{fit}|=3.05\times 10^{-5}$ \cite{PDG}; in the right panel of
Figure \ref{vtbdelta} we 
present a scan of values of the function of $J$ shown in Eq.~(\ref{Jfunct}),
for random values of $a$, $d$ and $\Delta_{12}$ (assuming $\Delta_{23}=0$). It is apparent that the
observed value of $J$ can be obtained quite generically when
$\Delta_{12}=\pi/2$, in a quite insensitive way to the specific
values of $a$ and $d$.

It is quite suggestive that some specific phase differences
between the elements $\ (H_u)_{23}\equiv f\ $ and $\ (H_d)_{23}\equiv
b\ $, and between  the elements $\ (H_u)_{12}\equiv y\ $ and $\ (H_d)_{12}\equiv
x\ $  and between the elements $\ (H_u)_{13}\equiv h\ $ and $\ (H_d)_{13}\equiv
g\ $, given by:
\bea
&&\Delta_{23}\ =\ \delta_f-\delta_b\ =\ 0\\
&&\Delta_{12}\ =\ \delta_y-\delta_x\ =\ \frac{\pi}{2}\\
&&\Delta_{13}\ =\ \delta_h-\delta_g\ =\ \frac{\pi}{2}
\eea
do lead to good generic values of both $V_{tb}$ and $J$. We are left
with four parameters $a,d$ and two combinations of
  phases independent of $\Delta_{12}$ and $\Delta_{23}$, all from the original mass matrices, which do
not seem to play any important role in obtaining a good
overall fit in the quark sector. 

Before we finish this subsection on the quark sector, we would like to
point out that since $a$ and $d$ can take almost any value (inside
their allowed range), one might actually get very close to a symmetric
limit, namely a $(2\leftrightarrow 3)$ family symmetry, relating
second and third families (see \cite{mutau} for implementations mostly
in the lepton sector).

Forgetting for a moment our ansatz but assuming the
$(2\leftrightarrow 3)$ limit for both up and down quark 
mass matrices, one is then guaranteed to have vanishing elements
$(W_u)_{13}$ and $(W_d)_{13}$ (i.e. we recover our ansatz). Moreover
some of the elements of the mass matrix are subject to the constraints
\bea
x=g\ \  &&\ \  y=h\\
\alpha=a \ \ && \ \  \beta=d\\
\delta_b\  = \ &\pi  & \  = \  \delta_f.
\eea
From these equations, we must have $\Delta_{23}=\delta_f-\delta_b=0$ and using Eq.~(\ref{Ceq}) it
is easy to see that $|V_{tb}|=1$ exactly, which means also that
$|V_{ub}|=|V_{cb}|=|V_{td}|=|V_{ts}|=0$. Since the fitted values of
$|V_{ub}|$, $|V_{cb}|$, $|V_{td}|$ and $|V_{ts}|$ are at most $10^{-2}$
and much smaller than $V_{tb}$, then it seems plausible that a small
deviation from this symmetric limit can easily restore the
experimental values of these $V_{ij}$'s.

To obtain a correct experimental fit, one would also have to require
the phases of $x$ and $y$ to be separated by $\pi/2$ (as remarked
earlier), and moreover the values of $\rho$ and $\gamma$ will have to
be chosen so as to obtain the correct ratios
$\frac{|V_{ub}|}{|V_{cb}|}$ and $\frac{|V_{td}|}{|V_{ts}|}$ (see
Eq.~(\ref{VubVcb})).

Although we imposed a flavor ansatz mainly for empirical reasons and simplicity,
it is interesting that the patterns emerging from it do actually lead
to a possible type of symmetry. Of course we just considered one of
the possible cases of our scheme, and others cases might lead to
different patterns or reveal some other feature and for this a more
systematic study is required, some of it being already underway.

\subsubsection{\bf The lepton sector}
In the lepton sector, the mixing matrix takes the same form as for the
quarks in the $(13-13)$ ansatz, i.e.
\bea
V&=& W_l^\dagger W_\nu= P_l^\dagger\ V_{{}_{PMNS}}
\eea
where $P_l$ is an unphysical diagonal phase matrix given below and
\bea
V_{{}_{PMNS}}=\pmatrix{
c_{\gamma'}c_{\rho'}+s_{\rho'}s_{\gamma'}|C'| e^{i\theta'}
&-c_{\gamma'}s_{\rho'}+s_{\gamma'}c_{\rho'}|C'| e^{i\theta'}
&s_{\gamma'} |S'|
\cr
-s_{\gamma'}c_{\rho'}+c_{\gamma'}s_{\rho'}|C'| e^{i\theta'}
& s_{\gamma'}s_{\rho'}+c_{\gamma'}c_{\rho'}|C'| e^{i\theta'}
& c_{\gamma'} |S'|
\cr
-s_{\rho'} |S'|
&-c_{\rho'}  |S'|
&
|C'|e^{-i\theta'} \cr }\ { P_\nu}
\eea
where $P_\nu$ is the Majorana phase matrix given below and the complex
rotation parameters $C',S', c_{\gamma'}$ and $c_{\rho'}$ are given by
\bea
C'&=& \frac{\left(\vphantom{\int_{\int}^{\int}}
  \sqrt{\displaystyle (m_\tau-a')(\lambda_3-d')}\ e^{i(\delta_f'-\delta_b')}  +\sqrt{(m_\tau-\alpha')(\lambda_3-\beta')}\right)
}{\sqrt{(2m_\tau-\alpha'-a')(2\lambda_3-\beta'-d')}}\ \ \ (i.e.\ \ |C'|\equiv
\ |V_{\tau 3}|)\ \  \label{vtau3}\\
S'&=&e^{i\delta_b'} \frac{\left(\vphantom{\int_{\int}^{\int}}
  \sqrt{\displaystyle (m_\tau-a')(\lambda_3-\beta')}- e^{i(\delta_f'-\delta_b')} \sqrt{(m_\tau-\alpha')(\lambda_3-d')}\right)
}{\sqrt{(2m_\tau-\alpha'-a')(2\lambda_3-\beta'-d')}} \\
c_{\gamma'}&=& \sqrt{\frac{m_\mu-\gamma'}{m_\mu-m_e}}\\
c_{\rho'} &=&\sqrt{\frac{\lambda_2-\rho'}{\lambda_2-\lambda_1}}
\eea
Note that $s_i^2 = 1-c^2_i$ and that $|C'|^2+|S'|^2=1$.
The CP phase $\theta'$ and the Majorana phase matrix $P_\nu$, as well as the
unphysical phase matrix $P_l$ are given respectively by
$\theta=\delta_{x'}-\delta_{y'}-\delta_{C'}\ $, $P_{\nu}=diag(1,e^{i\delta_{y'}},e^{i(\delta_{x'}+\delta_{C'})})\ $ and
$ P_l=diag(1,e^{i\delta_{x'}},e^{i(\delta_{y'}+\delta_{C'})})$, where $\ \delta_i\equiv \arg(i)$.

Again, the previous form of the mixing matrix implies the relations:
\bea
\frac{|V_{e3}|}{|V_{\mu
3}|}=\sqrt{\frac{\gamma'-m_e}{m_\mu-\gamma'}},\hspace{.7cm}
\frac{|V_{\tau  1}|}{|V_{\tau2}|}=\sqrt{\frac{\rho'-\lambda_1}{\lambda_2-\rho'}},\hspace{.7cm} \label{vtau1vtau2}
\eea
and
\bea
\frac{|V_{e3}|}{|V_{\tau 1}|}=\sqrt{\frac{(\gamma'-m_e)(\lambda_2-\lambda_1)}{(m_\mu-m_e)(\rho'-\lambda_1)}}, \hspace{.7cm}
\frac{|V_{\mu 3}|}{|V_{\tau 2}|}=\sqrt{\frac{(m_\mu-\gamma')(\lambda_2-\lambda_1)}{(m_\mu-m_e)(\lambda_2-\rho')}}.
\eea
with again only two parameters from the original mass
matrices separately controlling each ratio.

But, the lepton case is different, obviously, because the lepton mixing
matrix contains two large angles and also because the neutrino mass structure is
unknown and might not be as hierarchical as in the quark sector (see
for example \cite{neutrinoexp} for the latest global fits coming from
neutrino oscillation experiments). In the (13-13) ansatz we see that we must still enforce the charged lepton mass
matrix to have a very small first diagonal element, i.e
$\gamma'\equiv{\cal O}(m_e)$ in order to obtain a small value for
the ratio $|V_{e3}|/|V_{\mu3}|$.
This will imply that $a'+\alpha'=m_{\mu}+m_{\tau} + {\cal
  O}(m_e)$ (from the trace identity of $H_l$) which will simplify the
functional form of $V_{\tau3}$ of Eq.~(\ref{vtau3}). Before doing so, we need
to look also into the required value of $\rho'$ to obtain a correct
fit for $\frac{|V_{\tau 1}|}{|V_{\tau 2}|}\sim 1/\sqrt{2}$ in the
approximate tri-bimaximal scheme $(TBM)$
\cite{HarrisonPerkinsScott}. This leads to $\rho'\sim
\lambda_2+2\lambda_1$, and we can now use this relation to obtain the simplified form of $V_{\tau3}$:
\bea
 V_{\tau 3}(a',d',{\Delta'_{23}},\lambda_1)&=& \frac{\left(\vphantom{\int_{\int}^{\int}}
  \sqrt{\displaystyle (m_\tau-a')(D_{13} +\lambda_1-d')} e^{i{\Delta'_{23}}}+\sqrt{(a'-m_\mu)(\lambda_1+d')}\right)
}{\sqrt{(m_\tau-m_\mu)(D_{13}+2\lambda_1)}}\
\eea
where $D_{13}=(\lambda_3-\lambda_1)$ is fixed by the measured
atmospheric neutrino mass difference\footnote{Although we have been
writing $\lambda_1,\lambda_2$ and $\lambda_3$ for the neutrino eigenvalues, these could
in fact correspond to the squared physical masses if we consider the
neutrino mass matrix  squared as our starting point. In this case,
$D_{13}=({\Delta} m^2)_{atm}$.}.

We now have 7 free parameters (including 4 phases), since
 one of the neutrino masses is unknown, and we choose it to be $\lambda_1=m_1^2$.
\begin{figure}[t]
 \center
 \includegraphics[width=7.2cm,height=8.cm]{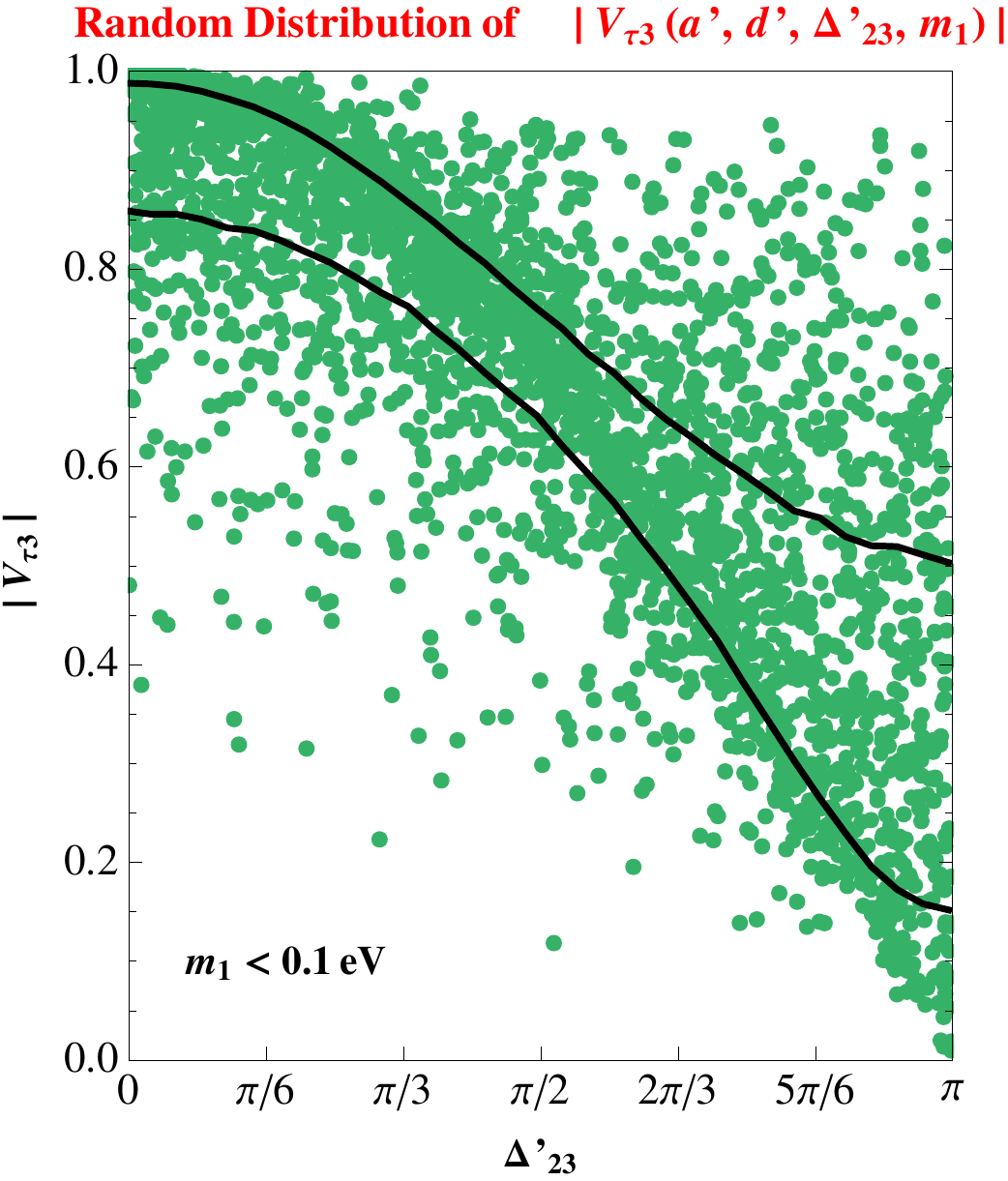}\hspace{1cm}
 \includegraphics[width=7.2cm,height=8.cm]{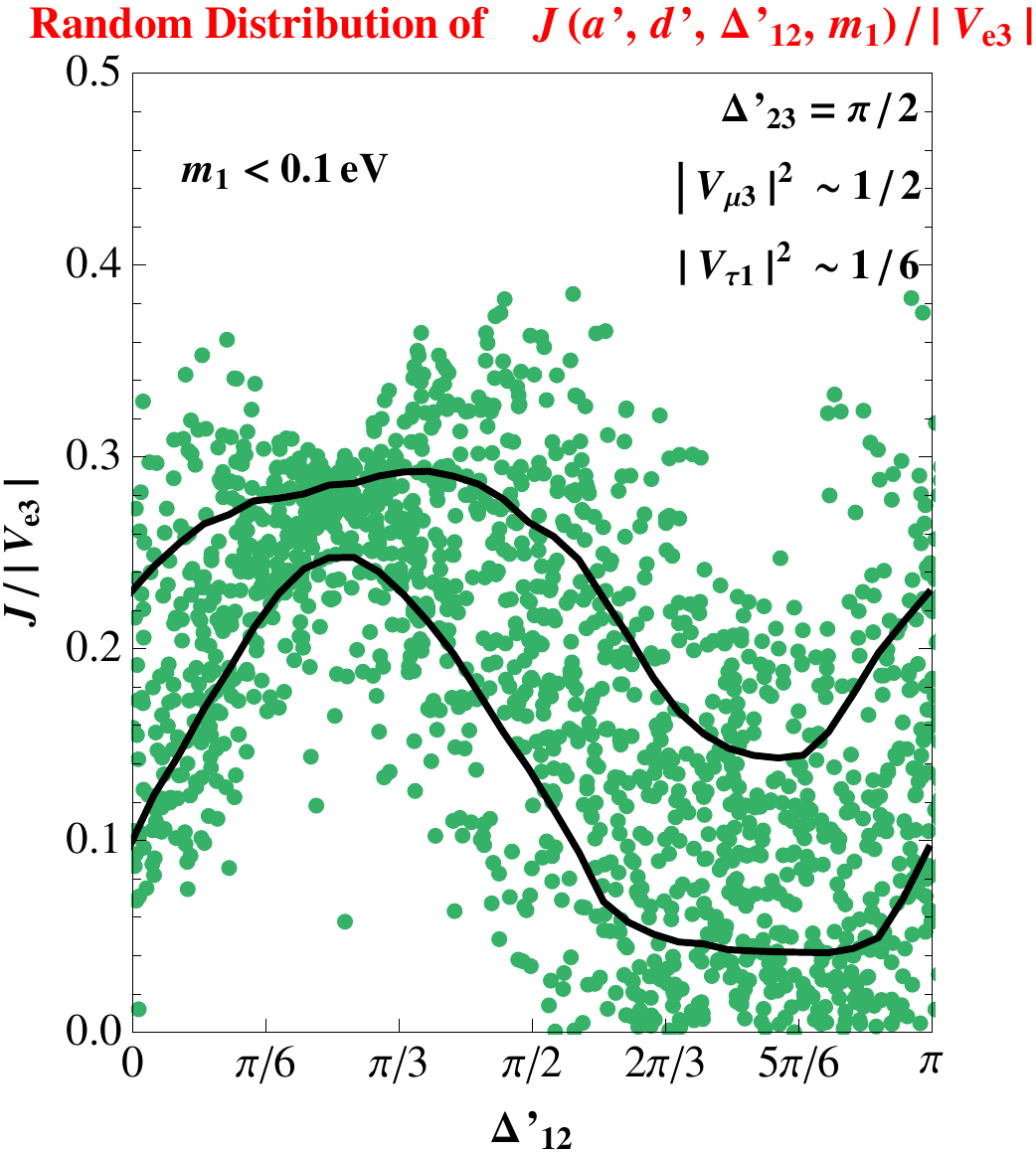}
 \caption{Distribution of $|V_{\tau3}(a',d',\Delta'_{23},m_1)|$ with random $a'$,
   $d'$, $\Delta'_{23}$ and $m_1$ (defined as the mass of the lightest neutrino) with respect to $\Delta'_{23}$
   (left panel). On the right panel, we present the
   distribution of the Jarlskog invariant $J(a',d',\Delta'_{12},m_1)$
   normalized to the value of $|V_{e3}|$, with
   random $a',d'$, $\Delta'_{12}$ and $m_1$, for fixed  $\Delta'_{23}=\pi/2$. On the two
   panels, the black curves represent the $25\%$ and $75\%$
   quantiles of the distribution for fixed $\Delta'_{23}$ (left) and
   $\Delta'_{12}$ (right). In other words $50\%$ of the random points
   lie between the curves, with $25\%$ above it and $25\%$ below it.}
\label{vtau3delta}
 \end{figure}
As seen in Figure \ref{vtau3delta}, when
$m_1<0.1$ eV, the correlation between the phase ${\Delta'_{23}}$ and the value of $|V_{\tau3}|$
is basically the same as it was in the quark sector, and
this is basically due to the fact that this feature happens for masses
which are hierarchical (and in fact, for larger values of $m_1$ the correlation between the phase
${\Delta'_{23}}$ and the value of $|V_{\tau3}|$ starts to wash out).
It is quite suggestive, that if $m_1<0.1$ eV and ${\Delta'_{23}}=\pi/2$, then
we obtain a generic size of $|V_{\tau 3}|\sim 1/\sqrt{2}$, as shown in
Figure \ref{vtau3delta}. Because $\gamma'$ is small, $c_{\gamma'}\sim
1$ and we will also have $|V_{\mu 3}| \sim 1/\sqrt{2}$.
These two values of $|V_{\mu 3}|$ and $|V_{\tau 3}|$ are consistent with
the tri-bimaximal scheme $(TBM)$ where $|V^{TBM}_{\tau 3}|= |V^{TBM}_{\mu 3}|= 1/\sqrt{2}$ and
so we find that in the lepton case the preferred value of the phase
combination $\Delta'_{23}\ $  is  $\ \frac{\pi}{2}$, which
statistically predicts a large value for both the mixing angles
$V_{\tau 3}$ and $V_{\mu 3}$.
It is also interesting to note that the charged fermion mass matrix
must have the same structure as the up and down quark mass matrices, namely that
the first diagonal element is of the order of the lightest eigenvalue.

Since the mixing element $V_{\tau1}$ is not small, we also conclude
that the neutrino mass structure must be different from the other
three matrices. This could be due to the fact that the eigenvalues are
not as hierarchical as for the charged fermions.
As in the quark case, it is interesting to note that further
imposing some flavor symmetry might simplify the relations obtained in
this ansatz. In particular 
one could ask what is the effect of imposing the $(2\leftrightarrow 3)$ family
symmetry (see for example \cite{mutau}) only in the neutrino sector.
In our case this is a natural question to ask since this symmetry imposes automatically the element
$(W_\nu)_{13}$ to be zero, which is the defining condition of our
ansatz. But it also imposes the following constraints on the mass
matrix elements 
\bea
(H_{\nu})_{12}=(H_{\nu})_{13},\ \ \  &i.e.&\ \  y'=h' \\
(H_{\nu})_{22}=(H_{\nu})_{33},\ \ \ &i.e.&\ \  \beta'=d' \\
\arg{(H_{\nu})_{23}}=\pi, \ \ \ \ \  &i.e.&\ \  \delta_{f'}=\pi
\eea
This will have interesting consequences on the lepton mixing matrix
because in this case we will have
\bea
 V_{\tau 3}\ =\ C'&=&\frac{1}{\sqrt{2}} \frac{\left(\vphantom{\int_{\int}^{\int}}
 \sqrt{(a'-m_\mu)} -\sqrt{\displaystyle (m_\tau-a')}\ e^{-i\delta_{b'}}  \right)
}{\sqrt{(m_\tau-m_\mu)}}\ +\ {\cal O}(m_e/m_\mu)
\eea
where we have assumed that $\gamma'\equiv {\cal O}(m_e)$. Now one can see that when $\delta_{b'}=\pi/2$, independent of anything
else, we will have $|V_{\tau 3}|=1/\sqrt{2}$, as well as
$|S'|=1/\sqrt{2}$. Thus the $2\leftrightarrow 3$ symmetry, along with the requirement of the
phase of $b'$ to be $\pi/2$, forces the appearance of large mixing
angle in the $V_{PMNS}$ matrix, and we remind the reader that we are
not considering a diagonal charged lepton mass matrix, although all our
matrices are taken in the $(13-13)$ ansatz.
The form of the $V_{PMNS}$ matrix under these assumptions is
\bea
V_{{}_{PMNS}}=\pmatrix{
c_{\gamma'}c_{\rho'}+s_{\rho'}s_{\gamma'} e^{i\theta'}/\sqrt{2}
&-c_{\gamma'}s_{\rho'} +s_{\gamma'}c_{\rho'}\ e^{i\theta'}/\sqrt{2}
&s_{\gamma'}/\sqrt{2}
\cr
-s_{\gamma'}c_{\rho'}+c_{\gamma'}s_{\rho'} e^{i\theta'}/\sqrt{2}
& s_{\gamma'}s_{\rho'}+c_{\gamma'}c_{\rho'}e^{i\theta'}/\sqrt{2}
& c_{\gamma'}/\sqrt{2}
\cr
-s_{\rho'}/\sqrt{2}
&-c_{\rho'}/\sqrt{2}
&
e^{-i\theta'}/\sqrt{2} \cr }\ P_\nu
\eea
where as before,
$c_{\gamma'}= \sqrt{\frac{m_\mu-\gamma'}{m_\mu-m_e}}\ $  and $\ c_{\rho'} =\sqrt{\frac{\lambda_2-\rho'}{\lambda_2-\lambda_1}} $
and where $\theta'$ is the Dirac CP phase given by
$\theta'=\delta_{x'}-\delta_{y'}-\delta_{C'}$ and the Majorana phase matrix
$P_\nu$ is given by
\bea
P_\nu=\pmatrix{
1&0&0\cr
0&e^{i\delta_{y'}}&0\cr
0&0&e^{i(\delta_{x'}+\delta_{C'})}}
\eea
The phase $\delta_{C'}$ is given by $\ \sin{\delta_{C'}}= \sqrt{\frac{m_\tau -a'}{m_\tau-m_\mu}}\ +\ {\cal O}(m_e/m_\mu)$.

Finally, in the more general case where one does not consider
$(2\leftrightarrow 3)$ symmetry in the
neutrino sector, the leptonic Jarlskog invariant $J$ is computed to be:
\bea
J&=& |V_{e3}||V_{\tau 1}||V_{\tau 3}|c_\rho' c_\gamma' \sin{\theta'}\\
&=& |V_{e3}||V_{\tau 1}|c_\rho' c_\gamma' {\frac{\left(\vphantom{\int_{\int}^{\int}}
  \sqrt{\scriptstyle  (m_{\tau}-a')(m_3-d')}\ \sin{\scriptstyle (\Delta'_{23}+\Delta'_{12})} +
  \sqrt{\scriptstyle (m_{\tau}-\alpha')(m_3-\beta')}\ \sin{\scriptstyle
    \Delta'_{12}}\right)}{\sqrt{\scriptstyle
      (2m_{\tau}-\alpha'-a')(2m_3-\beta'-d')}}}.
\eea
For the preferred value of $\Delta'_{23}=\pi/2$, such that
statistically it is more favorable to obtain a $V_{\tau3}$ mixing angle close to $1/\sqrt{2}$, we obtain
\bea
J&=& |V_{e3}||V_{\tau 1}|c_\rho' c_\gamma' {\frac{\left(\vphantom{\int_{\int}^{\int}}
  \sqrt{\scriptstyle  (m_{\tau}-a')(m_3-d')}\ \cos{\scriptstyle \Delta'_{12}} +
  \sqrt{\scriptstyle (m_{\tau}-\alpha')(m_3-\beta')}\ \sin{\scriptstyle
    \Delta'_{12}}\right)}{\sqrt{\scriptstyle
      (2m_{\tau}-\alpha'-a')(2m_3-\beta'-d')}}}.
\eea
This suggests that now, the value $\Delta'_{12}\sim\pi/4$ will maximize
the possible value of $J$ (normalized to $|V_{e3}|$), and this is confirmed in the random scan shown in Figure
\ref{vtau3delta}.

\section{Discussion and outlook}

In this work, we started by obtaining the simplest parametrization for the diagonalization
of a $3 \times 3$ hermitian matrix, as a function of both its matrix
elements and its eigenvalues.
Since the masses of fermions in the SM are well known, the problem to
attack is not an eigenvalue problem, but an eigenvector 
problem. In other words, we are able to obtain the eigenvectors of a
hermitian matrix in an algebraically compact form because we treated
the eigenvalues as known parameters, instead of unknown.
With this parametrization in hand, we wanted then to show how it
could simplify and make quite transparent the analysis and study of
some flavor schemes. To this end we defined a new flavor scheme which
imposes some (arguably obscure) constraints on the flavor structure in the fermion
sector. The constraint imposed is the requirement of the vanishing of
one of the mixing angles of the diagonalization matrix of each
hermitian fermion mass matrix of the SM. We called this the
``two-angle'' ansatz (i.e. out of three angles to diagonalize a
hermitian matrix, we consider the family of hermitian matrices
diagonalized by only two angles), and it is bottom-up
motivated, i.e. it is inspired on the observed structure of both
$V_{CKM}$ and $V_{PMNS}$ matrices. Nevertheless we also observe that in
flavor models where one requires a full symmetry between two of the
three families, like the $\mu-\tau$ symmetry models, one of the
consequences is precisely the vanishing of one of the mixing
angles.\footnote{In the case of a complex symmetric neutrino mass matrix, the
authors of \cite{Lavoura} studied the symmetry conditions required for the
vanishing of one of the mixing angles of the rotation matrix
}
 But we mainly
decided to keep the bottom-up approach motivation and study the
patterns emerging from our ``two-angle'' ansatz and focused on just
one possible implementation, the $(13-13)$ case, in
which the ${(W)}_{13}$ entries of all the diagonalizing matrices
happen to vanish.
Using our parametrization, one can actually write both $V_{CKM}$ and
$V_{PMNS}$ in terms of the original matrix elements and
eigenvalues. In particular we found a peculiar dependence of one of the
mixing angles with the model parameters. We observed that by fixing
the phase difference $\Delta_{23}$ between two parameters of the up
and down mass matrices (or the neutrino and charged lepton mass matrices), the mixing angle $V_{tb}$ (or
$V_{\tau3}$) can be ``predicted'' in a statistical sense, i.e. if one
makes a random scan allowing the remaining free parameters to take
any possible value, including the eigenvalues, one obtains a
narrow distribution for the value of $|V_{tb}|$ (or $|V_{\tau3}|$), and the central
value is a monotonic function of the phase difference $\Delta_{23}$.
This can be seen by rewriting the general formula for $V_{tb}$ in the $(13-13)$ ansatz
(see Eq.~(\ref{Ceq})) as
\bea
|V_{tb}|=F(\tilde{a},\tilde{\alpha},\tilde{d},\tilde{\beta},\Delta_{23})=Abs\left(\frac{\left(\vphantom{\int_{\int}^{\int}}
  \sqrt{\displaystyle (1-\tilde{a})(1-\tilde{d})}\ e^{i\Delta_{23}} +\sqrt{(1-\tilde{\alpha})(1-\tilde{\beta})}\right)
}{\sqrt{(2-\tilde{\alpha}-\tilde{a})(2-\tilde{\beta}-\tilde{d})}}\right)
\eea
where $\tilde{a}=a/m_t$, $\ \tilde{\alpha}=\alpha/m_t$,
$\ \tilde{d}=d/m_b$ and $\ \tilde{\beta}=\beta/m_b$, and
$a,\alpha$ and $d,\beta$ are diagonal elements of the up and down mass
matrices respectively.
If the mass eigenvalues $m_i\equiv\lambda^{u,d}_i$ are unconstrained, the only constraints on the
parameters required for a random scan are 
\bea
\lambda^u_1/\lambda^u_3\ <\ \tilde{\alpha},\tilde{a}\ <\ 1\hspace{1cm} &{\rm
and }&\hspace{1cm} \lambda^d_1/\lambda^d_3\ <
\ \tilde{\beta},\tilde{d}\ <\ 1
\eea
since the diagonal elements of a hermitian matrix must be bounded by
its largest and lowest eigenvalues.
The result of a scan over all these parameters (including the
eigenvalues) but for fixed $\Delta_{23} $ is a highly peaked distribution
centered at some value. This means that the generic value of $|V_{tb}|$ is controlled
almost exclusively by $\Delta_{23}$. Of course one may think that after
fixing the masses to the experimental values, as well as the other
mixing angles, maybe we might loose this statistical prediction. This
is not the case as was shown in Figure \ref{vtbdelta}, where the scan is
performed now with only two free parameters, the rest having been
fixed by other experimental observables.
There is still a clear correlation between the value of the phase
difference and the value of the angle $V_{tb}$.

The imposition of this specific two-angle ansatz (the ($13-13$) case) on
the fermion mass matrices amounts to 2 constraints per mass matrix,
and therefore 4 constraints in the quark sector. What we have noted is that with
only one more constraint, i.e. the fixing of the phase difference
$\Delta_{23}$, we are able to give a statistical prediction for the value of the
angle $|V_{tb}|$ (or $|V_{\tau3}|$), and this, irrespective of any other
parameter. In particular if $\Delta_{23}=0$ we would expect $|V_{tb}|\sim 1$ and in
the lepton sector, if we fix $\Delta'_{23}=\pi/2$, then we would expect
$|V_{\tau3}|\sim 1/\sqrt{2}$, both cases being close to the experimental
fits.
Two hermitian matrices contain 18 free parameters altogether and so it
is nontrivial that after imposing only 5 constraints on the whole set
we obtain one prediction, irrespective of anything else. 


Once noted this nontrivial property, we continued analyzing the rest of
consequences of our scheme and showed for example how the CP violating
phases in both quark and lepton sectors depend on the phases of the
original mass matrix elements. Another interesting outcome was the
realization of how to treat in a similar way the quarks and charged
lepton mass matrices, and use the special case of
the neutrino matrix to explain in a transparent way the
observed differences between the lepton and quark sectors. In
particular we also analyzed
the consequences of further imposing a $(2\leftrightarrow 3)$
symmetry in the neutrino sector.

It is true that out of many possible implementations of the two-angle
ansatz we chose to study only one case. We felt that this one case
would show the main features of the scheme in a transparent way, and so
we leave the systematic case by case study of the scenario for the future, although part
of this work is already underway.

\section{Acknowledgements}
M.T. would like to thank Alex Azatov, Joseph Schechter and Lijun Zhu for useful discussions.
C.H. wishes to thank Vestislav Apostolov, Fran\c cois Bergeron and Christophe Reutenauer for
many interesting discussions and helpful comments. This research was
partially funded by NSERC of Canada.

\section*{APPENDIX A: DIAGONALIZATION OF A HERMITIAN MATRIX}

We first note that the diagonal elements of a hermitian matrix are bounded by its
smallest and largest eigenvalue respectively. This means that in
order for a hermitian matrix to have zeroes in the diagonal
entries, we must have at least one eigenvalue positive and one
negative, the third one can take either sign, to accommodate the
zeroes in the diagonal elements. For simplicity, we do not consider this case
here and instead concentrate on positive definite hermitian
matrices such that the eigenvalues are all positive. Of course the
results can be trivially extended for the case of a more general
hermitian matrix, not necessarily definite positive.

Let's introduce our notation by considering the positive definite hermitian matrix $H$
\begin{eqnarray}
H=\pmatrix{ \gamma   & x   & g \cr x^* & \alpha   & b \cr g^* &
b^* & a \cr }. \label{Huapp}
\end{eqnarray}
\noindent The diagonal entries of $H$ must be real and positive and are bounded by
its smallest and largest eigenvalue respectively. Taking
$0<\lambda_1 < \lambda_2 < \lambda_3$ the bounds are
\bea
0<\lambda_1 \leq\ \gamma\ \leq \lambda_3\\
0<\lambda_1 \leq\ \alpha\ \leq \lambda_3 \\
0<\lambda_1 \leq\ a\ \leq \lambda_3
\eea
The off-diagonal entries of $H$ are also bounded by its smallest
and largest eigenvalues in the following way; since a $3 \times 3$
hermitian matrix has only three blocks of off-diagonal elements,
namely $(x,g)$, $(x,b^*)$ and $(g,b)$ up to conjugation \cite{KwongLiMathias}:
\bea
0 \leq \sqrt{|x|^2+|g|^2} \leq \frac{1}{2}\ |\lambda_3-\lambda_1|\\
0 \leq \sqrt{|g|^2+|b|^2} \leq \frac{1}{2}\ |\lambda_3-\lambda_1|\\
0 \leq \sqrt{|x|^2+|b|^2} \leq \frac{1}{2}\ |\lambda_3-\lambda_1|
\eea
which means that:
\bea
0 \leq |x| \leq \frac{1}{2}\ |\lambda_3-\lambda_1|\\
0 \leq |g| \leq \frac{1}{2}\ |\lambda_3-\lambda_1|\\
0 \leq |b| \leq \frac{1}{2}\ |\lambda_3-\lambda_1|
\eea
The above results are valid for any $N \times N$ hermitian matrix.
These bounds might not be very revealing in the quark and charged
lepton sectors, since the difference between the heaviest and the
lightest eigenvalues is of the order of the largest eigenvalue,
and so the constraint on the off-diagonal entries is quite mild.
On the other hand, if the difference between the heaviest and
lightest eigenvalue is very small then one sees that the
off-diagonal entries must actually be smaller than half that
difference, and so the hermitian matrix must be close to diagonal
form. This case might be possible in the neutrino sector in the
case of quasi-degenerate masses.
We now write the three invariants $Tr(H)$, $Tr(H^2)$ and $Det(H)$:
\begin{eqnarray}
Tr(H)&=&a+\alpha+\gamma=\lambda_1+\lambda_2+\lambda_3,  \\
Tr(H^2)&=&2(|x|^2+|b|^2+|g|^2)+a^2+\alpha^2+\gamma^2=\lambda_1^2+\lambda_2^2+\lambda_3^2  \\
Det(H)&=&\gamma(\alpha a -|b|^2)-a|x|^2 -\alpha |g|^2 + 2Re(bxg^*)= \lambda_1\lambda_2\lambda_3
\end{eqnarray}
They can be rewritten as
\begin{eqnarray}
\alpha&=&\lambda_1+\lambda_2+\lambda_3-a-\gamma,\\
|x|^2&=&\frac{(\gamma-\lambda_1)(\lambda_2-\gamma)(\lambda_3-\gamma)-|g|^2(\alpha-\gamma)+2 Re(bxg^*)}{(a-\gamma)}\\
|b|^2&=&\frac{(a-\lambda_1)(a-\lambda_2)(\lambda_3-a)+|g|^2(\alpha-a)-2 Re(bxg^*)  }{(a-\gamma)}
\end{eqnarray}
As noted in the text, the interesting thing of this notation is that the constraint formulae
on $x$ and $b$ actually become algebraic solutions for both $x$ and
$b$ when the term $Re(bxg^*)$ vanishes identically.

When one is interested in finding the eigenvalues of the
mass matrix given in Eq.~(\ref{Huapp}), it is necessary to find
solutions of the characteristic equation:
\begin{eqnarray}
0&=&(a-\lambda)((\lambda-\gamma)(\lambda-\alpha )-|x|^2)+(\lambda-\gamma) |b|^2 +\
(\lambda-\alpha ) |g|^2 + 2 Re (bxg^*) 
 \label{secular}
\end{eqnarray}
In our case, however, we do know the eigenvalues, which are
quantities measured experimentally. Therefore it is preferable to
treat them as known parameters instead of unknown variables. In this
case, we can obtain simple analytical forms for the unitary matrices
$W_u$ and $W_d$, responsible for diagonalizing the mass matrices
$H_u$ and $H_d$. Here is the simple procedure:
let $E_{u\lambda_i}\equiv(E^{\lambda_i}_1,E^{\lambda_i}_2,E^{\lambda_i}_3)$
be one eigenvector of the mass matrix $H_u$, i.e. $H_u\
E_{u\lambda_i}=\lambda_i\ E_{u\lambda_i}$, where $\lambda_i$ is one
of the eigenvalues of $H_u$. This means that $H_{u \lambda_i}\
E_{u\lambda_i} =0$, or simply 
\bea
Det(H_{u\lambda_i})=0\label{homogeneous}
\eea
which is in fact the characteristic equation of
$H_u$ given in Eq.~(\ref{secular}) and where 
\bea H_{u\lambda_i}=\pmatrix{
(\gamma-\lambda_i)   & x   & g \cr x^* & (\alpha-\lambda_i)   & b \cr
g^*   & b^* & (a-\lambda_i) \cr }
\eea 
Because, the $Det(H_{u\lambda_i})$ vanishes, we know that one row of the matrix must
be a linear combination of the other two. Depending on which row we
choose to treat as linearly dependent, we can obtain different (but
equivalent) parametrizations for the eigenvectors $E_{u\lambda_i}$.
Since the homogeneous equation~(\ref{homogeneous}) can be multiplied
by an arbitrary number, we will obtain ratios (of $2\times 2$
determinants) for the eigenvector components:

Using the first row as linearly dependent, we obtain:
\bea
E^{\lambda_i}_2/E^{\lambda_i}_1&=&-{ \left| \begin{array}{cc} x^* &
b  \\ g^* & (a-{\lambda_i}) \\  \end{array}\right| \over
\ \left| \begin{array}{cc} (\alpha - {\lambda_i}) & b  \\ b^* & (a-{\lambda_i}) \\  \end{array} \right|\ } \\
 E^{\lambda_i}_3/E^{\lambda_i}_1&=&{
\left| \begin{array}{cc} x^* & (\alpha-{\lambda_i})  \\ g^* & b^* \\
\end{array}\right| \over \ \left| \begin{array}{cc} (\alpha -
{\lambda_i}) & b  \\ b^* & (a-{\lambda_i}) \\  \end{array} \right|\
} \eea To simplify the notation, we can choose
$E^{\lambda_i}_1\equiv\left| \begin{array}{cc} \alpha - {\lambda_i} & b
\\ b^* & a-{\lambda_i}
    \\  \end{array}\right|$
and obtain finally the eigenvector 
\bea E^{1st\ row}_{u{\lambda_i}
}\equiv \left(\left| \begin{array}{cc} (\alpha - {\lambda_i}) & b
\\ b^* & (a-{\lambda_i})  \\  \end{array}\right|, -\left|
\begin{array}{cc} x^* & b  \\ g^* & (a-{\lambda_i}) \\
\end{array}\right|,
\left| \begin{array}{cc} x^* & (\alpha-{\lambda_i})  \\ g^* & b^* \\
  \end{array}\right|\right) \label{E1row}
\eea
If instead, we choose row 2 as linearly dependent we obtain 
\bea E^{2nd\
row}_{u{\lambda_i}}\equiv \left(\left| \begin{array}{cc} x & g  \\
b^* & (a-{\lambda_i})  \\  \end{array}\right|, -\left|
\begin{array}{cc} (\gamma-\lambda_i) & g  \\ g^* & (a-{\lambda_i})
\\  \end{array}\right|,
\left| \begin{array}{cc} (\gamma -\lambda_i) & x  \\ g^* & b^* \\
  \end{array}\right|\right)\label{E2row}
\eea
and finally when the third row is treated as linearly dependent we
have 
\bea E^{3d\ row}_{u{\lambda_i}}\equiv \left(\left|
\begin{array}{cc} x & g  \\  (\alpha-{\lambda_i})&b  \\
\end{array}\right|, -\left| \begin{array}{cc} (\gamma-\lambda_i) & g
\\ x^* & b \\  \end{array}\right|,
\left| \begin{array}{cc} (\gamma-\lambda_i) & x  \\ x^* &(\alpha-\lambda_i) \\
  \end{array}\right|\right)\label{E3row}
\eea
Since there are three eigenvalues, and three parametrization choices
for each, we have 9 equivalent parametrizations of the unitary
matrix $W_u$ constructed using any of the previous eigenvectors. Note that
the difference between each parametrization, is the explicit absence
of one of the diagonal elements in the formulae ($\gamma$ in the
first one, $\alpha$ in the second one and $a$ in the third one). Of
course, the characteristic equation~(\ref{secular}) relates each
'basis' with each other. 
In the text, we used this parametrization freedom to
choose a mixing matrix $W$ that has the correct form when $b\to
0,\ g\to 0$ and $x\to 0$. In that limit, the different 
parametrization can be written as: 
\bea
E^{1st\ row}_{u\lambda_i}&\equiv& \left((\alpha-\lambda_i)(a-\lambda_i),0,0\right)\\
E^{2nd\ row}_{u\lambda_i}&\equiv& \left(0,-(\gamma-\lambda_i)(a-\lambda_i),0\right)\\
E^{3d\ row}_{u\lambda_i}&\equiv&
\left(0,0,(\alpha-\lambda_i)(\gamma-\lambda_i)\right) 
\eea 
Some of these, can become problematic since in this limit, we have
$\ (\gamma-\lambda_1)\to 0$, $\ (\alpha-\lambda_2)\to 0\ $ and
$\ (a-\lambda_3)\to 0\ $. There is however, out of the 9 possible
combinations, one single choice which has the correct smooth
asymptotic behavior. The choice is to take $E^{1st\
row}_{u\lambda_1}$ for $\lambda_1$, $\ E^{2nd\ row}_{u\lambda_2}$ for
$\lambda_2$ and $\ E^{3d\ row}_{u\lambda_3}$ for $\lambda_3$, i.e.
\begin{equation}
\hspace{-.3cm}
W= \left(\begin{array}{ccc}
\frac{\displaystyle(\alpha-\lambda_1)(a-\lambda_1)-|b|^2}{\displaystyle N_1}
& \frac{\displaystyle gb^*-x(a-\lambda_2)}{\displaystyle N_2}
& \frac{\displaystyle \displaystyle xb -g(\alpha-\lambda_3)}{\displaystyle N_3}\\
\frac{\displaystyle g^*b-x^*(a-\lambda_1)}{\displaystyle N_1}
& \frac{\displaystyle (\gamma-\lambda_2)(a-\lambda_2)-|g|^2}{\displaystyle N_2}
& \frac{\displaystyle x^*g-b(\gamma-\lambda_3)}{\displaystyle N_3}\\
\frac{\displaystyle x^*b^*-g^*(\alpha-\lambda_1)}{\displaystyle N_1}
& \frac{\displaystyle xg^*-b^*(\gamma-\lambda_2)}{\displaystyle N_2}
& \frac{\displaystyle (\gamma-\lambda_3)(\alpha-\lambda_3)-|x|^2}{\displaystyle N_3}
\end{array}\right)
\end{equation}
with the normalization parameters
\begin{eqnarray}
N_1^2&=&(\lambda_3-\lambda_1)(\lambda_2-\lambda_1) \left[(\alpha-\lambda_1)(a-\lambda_1)-|b|^2\right],    \\
N_2^2&=&(\lambda_3-\lambda_2)(\lambda_2-\lambda_1) \left[(a-\lambda_2)(\lambda_2-\gamma)+|g|^2\right],    \\
N_3^2&=&(\lambda_3-\lambda_2)(\lambda_3-\lambda_1) \left[(\lambda_3-\gamma)(\lambda_3-\alpha)-|x|^2\right]
\end{eqnarray}
Of course depending on the specific scenario studied, some other
parametrization might be used. For example, one can use only the
eigenvectors of Eq.~(\ref{E1row}), i.e.
\begin{equation}
\hspace{-.3cm}
W= \left(\begin{array}{ccc}
\frac{\displaystyle(\alpha-\lambda_1)(a-\lambda_1)-|b|^2}{\displaystyle N_1}
&\frac{\displaystyle(\alpha-\lambda_2)(a-\lambda_2)-|b|^2}{\displaystyle N'_2}
&\frac{\displaystyle(\alpha-\lambda_3)(a-\lambda_3)-|b|^2}{\displaystyle N'_3}\\
\frac{\displaystyle g^*b-x^*(a-\lambda_1)}{\displaystyle N_1}
&\frac{\displaystyle g^*b-x^*(a-\lambda_2)}{\displaystyle N'_2}
&\frac{\displaystyle g^*b-x^*(a-\lambda_3)}{\displaystyle N'_3}\\
\frac{\displaystyle x^*b^*-g^*(\alpha-\lambda_1)}{\displaystyle N_1}
&\frac{\displaystyle x^*b^*-g^*(\alpha-\lambda_2)}{\displaystyle N'_2}
&\frac{\displaystyle x^*b^*-g^*(\alpha-\lambda_3)}{\displaystyle N'_3}\\
\end{array}\right), \label{E1rowpar}
\end{equation}
or use the ones from Eq.~(\ref{E2row}):
\begin{equation}
\hspace{-.3cm}
W= \left(\begin{array}{ccc}
 \frac{\displaystyle gb^*-x(a-\lambda_1)}{\displaystyle N''_1}
& \frac{\displaystyle gb^*-x(a-\lambda_2)}{\displaystyle N_2}
& \frac{\displaystyle gb^*-x(a-\lambda_3)}{\displaystyle N''_3}\\
 \frac{\displaystyle (\gamma-\lambda_1)(a-\lambda_1)-|g|^2}{\displaystyle N''_1}
& \frac{\displaystyle (\gamma-\lambda_2)(a-\lambda_2)-|g|^2}{\displaystyle N_2}
& \frac{\displaystyle (\gamma-\lambda_3)(a-\lambda_3)-|g|^2}{\displaystyle N''_3}\\
 \frac{\displaystyle xg^*-b^*(\gamma-\lambda_1)}{\displaystyle N''_1}
& \frac{\displaystyle xg^*-b^*(\gamma-\lambda_2)}{\displaystyle N_2}
& \frac{\displaystyle xg^*-b^*(\gamma-\lambda_3)}{\displaystyle N''_3}
\end{array}\right), \label{E2rowpar}
\end{equation}
or use the ones from Eq.~(\ref{E3row}):
\begin{equation}
\hspace{-.3cm}
W= \left(\begin{array}{ccc}
  \frac{\displaystyle \displaystyle xb -g(\alpha-\lambda_1)}{\displaystyle N'''_1}
& \frac{\displaystyle \displaystyle xb -g(\alpha-\lambda_2)}{\displaystyle N'''_2}
& \frac{\displaystyle \displaystyle xb -g(\alpha-\lambda_3)}{\displaystyle N_3}\\
  \frac{\displaystyle x^*g-b(\gamma-\lambda_1)}{\displaystyle N'''_1}
& \frac{\displaystyle x^*g-b(\gamma-\lambda_2)}{\displaystyle N'''_2}
& \frac{\displaystyle x^*g-b(\gamma-\lambda_3)}{\displaystyle N_3}\\
  \frac{\displaystyle (\gamma-\lambda_1)(\alpha-\lambda_1)-|x|^2}{\displaystyle N'''_1}
& \frac{\displaystyle (\gamma-\lambda_2)(\alpha-\lambda_2)-|x|^2}{\displaystyle N'''_2}
& \frac{\displaystyle (\gamma-\lambda_3)(\alpha-\lambda_3)-|x|^2}{\displaystyle N_3}
\end{array}\right), \label{E3rowpar}
\end{equation}
where the normalization constants $N''_1$, $N'''_1$, $N'_2$,
$N'''_2$, $N'_3$ and $N''_3$ can easily be obtained from $N_1$, $N_2$ and
$N_3$ by permutation of the mass eigenvalues.

All in all, one can write the matrix $W$ with 9 different
parametrizations (modulo a diagonal phase matrix), by permutations of the column vectors from
Eqs.~(\ref{E1row}), (\ref{E2row}) and (\ref{E3row}).
Of course, when one takes some special limits some of the
parametrizations will reveal themselves less useful, as it is possible
to find undetermined expressions of the type $\ 0/0$. For example in the
parametrization shown in Eq.~(\ref{E2rowpar}) this will happen if we
take simultaneously the limits $\ g\to 0$ and $\ x\to 0 \ $. In that situation one simply
chooses the parametrization with a smooth limit.

It is easy to realize from Eqs.~(\ref{E1rowpar}), (\ref{E2rowpar}) and (\ref{E3rowpar})
that one can obtain very simple expressions for the absolute value of each
element of W. By taking the real elements of these three parametrizations
and squaring them we obtain, in terms of the eigenvalues and the parameters
$\gamma$, $a$, $x$ and $g$:
\begin{eqnarray}
|W_{11}|^2 &=&\frac{(\lambda_2-\gamma)(\lambda_3-\gamma)
+|x|^2+|g|^2}{(\lambda_3-\lambda_1)(\lambda_2-\lambda_1)}
 \end{eqnarray}
 \begin{eqnarray}
|W_{12}|^2 &=&\frac{(\gamma-\lambda_1)(\lambda_3-\gamma)
-|x|^2-|g|^2}{(\lambda_3-\lambda_2)(\lambda_2-\lambda_1)}
 \end{eqnarray}
 \begin{eqnarray}
|W_{13}|^2 &=&\frac{|x|^2+|g|^2-(\gamma-\lambda_1)(\lambda_2-\gamma)
}{(\lambda_3-\lambda_2)(\lambda_3-\lambda_1)}
\end{eqnarray}
\begin{eqnarray}
|W_{21}|^2 &=&\frac{(\gamma-\lambda_1)(a-\lambda_1)
-|g|^2}{(\lambda_3-\lambda_1)(\lambda_2-\lambda_1)}
 \end{eqnarray}
 \begin{eqnarray}
|W_{22}|^2 &=&\frac{|g|^2 +(\lambda_2-\gamma)(a-\lambda_2)
}{(\lambda_3-\lambda_2)(\lambda_2-\lambda_1)}
 \end{eqnarray}
 \begin{eqnarray}
|W_{23}|^2 &=&\frac{(\lambda_3-\gamma)(\lambda_3-a)
-|g|^2}{(\lambda_3-\lambda_2)(\lambda_3-\lambda_1)}
\end{eqnarray}
\begin{eqnarray}
|W_{31}|^2 =\frac{(\gamma-\lambda_1)(\lambda_3+\lambda_2-\gamma-a)-|x|^2
}{(\lambda_3-\lambda_1)(\lambda_2-\lambda_1)}
\end{eqnarray}
\begin{eqnarray}
|W_{32}|^2 =\frac{(\lambda_2-\gamma)(\lambda_3+\lambda_1-\gamma-a)+|x|^2
}{(\lambda_3-\lambda_2)(\lambda_2-\lambda_1)}
\end{eqnarray}
\begin{eqnarray}
|W_{33}|^2 =\frac{(\lambda_3-\gamma)(\gamma+a-\lambda_1-\lambda_2)-|x|^2
}{(\lambda_3-\lambda_2)(\lambda_3-\lambda_1)}
\end{eqnarray}


\section*{APPENDIX B: THE $\ (13-13)\ $ ANSATZ}

Let's recall the notation for $H_u$ and $H_d$:
\bea
H_u=\pmatrix{ \gamma   & x   & g \cr x^* & \alpha   & b \cr g^* &
b^* & a \cr }
,\phantom{pp}
 H_d=\pmatrix{ \rho  & y   & h \cr y^* &
 \beta   & f \cr h^*  & f^* & d \cr} .
\eea
$W_u$ and $W_d$ are the unitary transformations diagonalizing $H_u$ and $H_d$ respectively.
We want to find the parametrization of both $W_u$ and $W_d$ when the
elements $(W_u)_{13}$ and $(W_d)_{13}$ vanish.
The requirement for the cancellation of these elements is 
$\ \ xb-g(\alpha-m_t)\!=\!0\ \ $ and  $\ \ yf-h(\beta-m_b)\!=\!0\ \ $,
and with them we obtain in both sectors the simpler identities:
\begin{eqnarray}
|x|^2=\frac{(\gamma-m_u)(m_c-\gamma)(m_t-\alpha)}{(2m_t-\alpha -a)}\hspace{2cm}&&
|y|^2=\frac{(\rho-m_d)(m_s-\rho)(m_b-\beta)}{(2m_b-\beta-d)}\nonumber\\
|g|^2=\frac{(\gamma-m_u)(m_c-\gamma)(m_t-a)}{(2m_t-\alpha -a)}\hspace{2cm}&&
|h|^2=\frac{(\rho-m_d)(m_s-\rho)(m_b-d)}{(2m_b-\beta-d)}\nonumber\\
|b|^2=(m_t-\alpha)(m_t-a)\hspace{3.7cm}&&
|f|^2=(m_b-\beta)(m_b-d)\nonumber\\
\delta_{x} -\delta_{g} + \delta_{b}=\pi\hspace{3.7cm}&&\hspace{1cm}
\delta_{y} -\delta_{h} + \delta_{f}=\pi
\end{eqnarray}
From the above expressions we obtain the following simple
parametrization of up and down quark mixing matrices, with
$(W_u)_{13}=0$, $(W_d)_{13}=0$ :
\begin{eqnarray}
W_u^{\dagger}\!=\!\pmatrix{ \sqrt{\frac{(m_c-\gamma)}{(m_c-m_u)}} &
-\sqrt{\frac{(\gamma-m_u)(m_t-\alpha)}{(m_c-m_u)(2m_t-\alpha-a)}}e^{i\delta_{x}}
 & \sqrt{\frac{(\gamma-m_u)(m_t-a)}{(m_c-m_u)(2m_t-\alpha -a)}}e^{i(\delta_{x}+\delta_{b})} \cr
-\sqrt{\frac{(\gamma-m_u)}{(m_c-m_u)}}e^{-i\delta_{x}} &
-\sqrt{\frac{(m_c-\gamma)(m_t-\alpha)}{(m_c-m_u)(2m_t-\alpha -a)}}
&
\sqrt{\frac{(m_c-\gamma)(m_t-a)}{(m_c-m_u)(2m_t-\alpha -a)}}e^{i\delta_{b}}\cr
0 & \sqrt{\frac{(m_t-a)}{(2m_t-\alpha -a)}}e^{-i\delta_{b}} &
\sqrt{\frac{(m_t-\alpha)}{(2m_t-\alpha-a)}} \cr} \ \
\end{eqnarray}
and
\begin{eqnarray}
\hspace{-.3cm}W_d\!=\!\pmatrix{ \sqrt{\frac{(m_s-\rho)}{(m_s-m_d)}} &
-\sqrt{\frac{(\rho-m_d)}{(m_s-m_d)}}e^{i\delta_{y}} & 0 \cr
-\sqrt{\frac{(\rho-m_d)(m_b-\beta)}{(m_s-m_d)(2m_b-\beta-d)}}e^{-i\delta_{y}}
&
-\sqrt{\frac{(m_s-\rho)(m_b-\beta)}{(m_s-m_d)(2m_b-\beta -d)}}
& \sqrt{\frac{(m_b-d)}{(2m_b-\beta-d)}}e^{i\delta_{f}}\cr
\sqrt{\frac{(\rho-m_d)(m_b-d)}{(m_s-m_d)(2m_b-\beta -d)}}e^{-i(\delta_{y}+\delta_{f})}
&
\sqrt{\frac{(m_s-\rho)(m_b-d)}{(m_s-m_d)(2m_b-\beta-d)}}e^{-i\delta_{f}}
& \sqrt{\frac{(m_b-\beta)}{(2m_b-\beta -d)}} \cr} .\ \
\end{eqnarray}
One can now check that in this ansatz,
which basically shows that the structure of the quark mixing matrix is
\bea
V_{{}_{CKM}}=\pmatrix{
c_{\gamma}c_\rho+s_{\rho}s_{\gamma} e^{i(\delta_x-\delta_y)} C^*
&-c_{\gamma}s_\rho\ e^{i\delta_y}+s_{\gamma}c_{\rho}\ e^{i\delta_x} C^*
&s_\gamma\ S\ e^{i\delta_x}
\cr
-s_{\gamma}c_\rho\ e^{-i\delta_x}+c_{\gamma}s_{\rho} e^{-i\delta_y} C^*
& s_{\gamma}s_{\rho}\ e^{i(\delta_y-\delta_x)}+c_{\gamma}c_{\rho}C^*
& c_\gamma\ S
\cr
-s_{\rho}\ S^*  e^{-i\delta_y}
&-c_{\rho}\  S^*
&
C
\cr } 
\eea 
where $C$, $S$, $c_\gamma$ and $c_\rho$ are given by
\bea C&=&
\frac{\left(\vphantom{\int_{\int}^{\int}}
  \sqrt{\displaystyle (m_t-a)(m_b-d)}\ e^{i(\delta_f-\delta_b)} +\sqrt{(m_t-\alpha)(m_b-\beta)}\right)
}{\sqrt{(2m_t-\alpha-a)(2m_b-\beta-d)}}\ \equiv\ V_{tb} \\
S&=&e^{i \delta_b} \frac{\left(\vphantom{\int_{\int}^{\int}}
  \sqrt{\displaystyle (m_t-a)(m_b-\beta)}- e^{i(\delta_f-\delta_b)} \sqrt{(m_t-\alpha)(m_b-d)}\right)
}{\sqrt{(2m_t-\alpha-a)(2m_b-\beta-d)}}\\
c_{\gamma}&=& \sqrt{\frac{m_c-\gamma}{m_c-m_u}}\\
c_\rho &=&\sqrt{\frac{m_s-\rho}{m_s-m_d}}
\eea
and $s_i=\sqrt{1-c_i^2}$ and $|C|^2+|S|^2=1$. The previous form for $V_{CKM}$ can then be put in the form given in
Eq.(\ref{vckm}) by pulling out two diagonal phase matrices $P_u$ and
$P_d$ as given in the main text.

We can also quickly compute the Jarlskog invariant $J$ for this mixing
matrix, for example by computing
\bea
J&=&{\rm Im}(V_{cb}V^*_{tb}V^*_{cd}V_{td})\non\\
J&=&{\rm Im}\left(-c_\gamma^2s_\rho^2|SC|^2 + c_\gamma c_\rho s_\gamma s_\rho
|S|^2 e^{i(\delta_x-\delta_y)} C^*\right)\non\\
J&=& c_\gamma c_\rho s_\gamma s_\rho |S|^2 {\frac{\left(\vphantom{\int_{\int}^{\int}}
  \sqrt{\scriptstyle  (m_t-a)(m_b-d)}\ \sin{\scriptstyle (\delta_b\!-\!\delta_f\!+\!\delta_x\!-\!\delta_y)} +
  \sqrt{\scriptstyle (m_t-\alpha)(m_b-\beta)}\ \sin{\scriptstyle
    (\delta_x\!-\!\delta_y)}\right)}{\sqrt{\scriptstyle
      (2m_t-\alpha-a)(2m_b-\beta-d)}}}\
\eea

\end{document}